THE AMERICAN UNIVERSITY IN CAIRO

SCHOOL OF SCIENCE AND ENGINEERING

# EVIDENCE FOR QUASI-PERIODIC OSCILLATIONS IN THE RECURRENT EMISSION FROM MAGNETARS AND THEIR IMPLICATIONS ON THE NEUTRON STAR PROPERTIES AND EQUATION OF STATE

A Thesis Submitted to
The Department of Physics
In Partial Fulfillment of the Requirements for the degree of
Master of Science

In the subject of
Physics

By

Ahmed Monzer El Mezeini

2010

The American University in Cairo

Thesis Title: Evidence for Quasi-Periodic Oscillations in the Recurrent Emission from Magnetars and their implications on the Neutron Star Properties and Equation of State

A Thesis Submitted by Ahmed Monzer El Mezeini

To the Department of Physics

May 2010

In partial fulfillment of the requirements for

The degree of Master of Science

# Acknowledgements

I met Dr. Alaa Ibrahim in March 2008, where he introduced me to the exciting world of High-Energy Astrophysics and how it uncovers connections between particle physics, nuclear physics, gravitation and states of matter. I was instantly hooked and I decided that I would do something similar. My passion still lies there, and I have never looked back. His contagious enthusiasm has such a magical effect on me that discussing physics with him truly constitutes an exciting part of my life. I appreciate his guidance and mentorship during which I had to learn a lot of new topics in the domain of extreme physics. He kindly shared with me his experience and knowledge and offered helpful advice on the transition from a student to a young researcher.

I am very grateful to Dr. Amr Shaarawi, who has been an exceptional mentor throughout my graduate studies. I have truly benefited from our numerous discussions and his guidance on topics that I was trying to learn. His view is always deep and broad with a unique insight. His influence has significantly affected the way I see and approach problems.

I was very fortunate to be among the members of the Physics department at AUC, which is full of great minds and personalities. I am indebted to Dr. Salah El Nahwy for constantly advising me on how to become a rigorous physicist, Dr. Salah El Sheikh for the continuous career guidance and sharing his personal experience, and Dr. Ahmed Aboul Seoud for the vibrant discussions, which taught me a number of topics in theoretical physics and mathematics and for the advice on future research. I am thankful for Dr. Sherif Sedky and Dr. Ehab Abdel Rahman for giving me the opportunity to work on a cutting-edge research project in Japan under the supervision of one of the world's renowned experts in semiconductor physics Dr. Fujio Masuoka. I thank Dr. Hosny Omar for constantly encouraging and supporting my research work.

Finally, I wish to express my sincerest appreciation and gratitude to my parents 'Monzer and Fatena', my sisters 'Reem and Reham' and my partner 'Heba' for their unconditional support and encouragement. They always supported my decisions and motivated me to work harder to reach the peak of my aspirations. From them, I have learned that what matters is the journey itself rather than arriving at the destination. Although my father did not get to see this manuscript in its final form because his life's journey was recently cut short by his courageous struggle with a heart condition, I dedicate this manuscript to the memory of my father, a loving and generous human being and an amazing parent.

My research work at the Kavli Institute for Astrophysics and Space Research at the Massachusetts Institute of Technology was generously supported by American University in Cairo Research Grant. This is largely due to the generous support from Dr. Hosny Omar (Chair of the Department of Physics), Dr. Medhat Haroun (Dean of the School of Science and Engineering), and Dr. Ali Hadi (Vice Provost and Director of Graduate Studies and Research).

# Abstract

In this thesis, we present an analysis of highly magnetized neutron stars "magnetars", in search for high frequency oscillations in the recurrent emission from the soft gamma repeater SGR 1806-20, and we discuss the physical interpretation of these oscillations and its implications on the neutron star properties and structure. We present evidence for Quasi-Periodic Oscillations (QPOs) in the recurrent outburst activity from the soft gamma repeater SGR 1806-20 using Rossi X-ray Timing Explorer (RXTE) observations.

By searching a large sample of bursts for timing signals at the frequencies of the QPOs discovered in the 2004 December 27 giant flare from the source, we find three QPOs at 84, 103, and 648 Hz in three different bursts. The first two QPOs lie within 8.85% and 11.83%, respectively, from the 92 Hz QPO detected in the giant flare. The third QPO lie within 3.75% from the 625 Hz QPO also detected in the same flare. These QPOs are detected in archival observations that took place eight years before the giant flare. The detected QPOs are found in bursts with different durations, morphologies, and brightness, and are vindicated by Monte Carlo simulations. We also find evidence for candidate QPOs at higher frequencies (1095, 1230, 2785 and 3690 Hz) in other bursts with lower statistical significance. The fact that we can find evidence for QPOs in the recurrent bursts at frequencies relatively close to those found in the giant flare is intriguing and can offer insight about the origin of the oscillations.

We confront our findings against the available theoretical models and discuss the physical interpretation of these QPOs. The leading interpretation for the origin of magnetar QPOs suggests that these toroidal seismic oscillatory modes are most likely to be excited by a magnetar crustquake of the neutron star crust. Other models have been proposed to explain the QPO phenomena we observe in magnetars including magnetospheric oscillations, magnetic flux tubes and modes of a passive debris disk.

After understanding the real nature of quasi-periodic X-ray oscillations, their observation will be very useful to put stringent constraints on neutron star masses and radii, which define the neutron star equation of state (EOS). We discuss the connection between the QPOs we report and those detected in the giant flares as well as their implications to the neutron star properties.

*Dedicated to my late father Monzer El Mezeini,*

*my mother Fatena El Mezeini,*

*my sisters Reem and Reham,*

*and my partner Heba Gabr*

*I did not give up, because you were there for me.*

*Thank you for everything!*

# Contents

# List of Figures

# List of Tables

# Chapter 1

# Introduction

## 1.1 Neutron Stars: An Overview

### 1.1.1 Introduction

In this chapter we provide an overview of this exciting field of high-energy astrophysics. We discuss the history of the discovery of neutron stars, the birth of highly magnetized neutron stars and their connection to supernova, the main properties and the structure of the neutron star are summarized, followed by a discussion of the type of object we are investigating in this thesis, which are Soft Gamma Repeaters (SGRs) and finally we conclude with a quick synopsis of the Rossi X-ray timing Explorer space satellite, which we used its observational data for our work.

### 1.1.2 Prediction and discovery

In the year 1932, James Chadwick announced his discovery of the 'neutron', the neutral particle that exists within the atomic nucleus alongside with the protons. One year later, the soviet physicist Lev Landau predicted the existence of neutron stars. Baade and Zwicky (1934) devised the theoretical conception of neutron stars and proposed their connection to supernova. Since then, neutron stars have been considered to be extremely fascinating celestial compact objects with exotic properties. In 1967, Baade and Zwicky's hypothesis proved to be true and the existence of neutron stars was confirmed due to the work of Jocelyn Bell a graduate student at the University of Cambridge at the time working under the supervision of Antony Hewish (Hewish et al. 1968). Hewish and Bell detected a signal from a radio pulsar that is spinning several times per second; this signal had the characteristic feature of having regular pulsations, which distinguished it from typical background scintillation signals. Hewish was awarded the Nobel Prize in Physics in 1974 for the discovery of pulsars, which confirmed the existence of neutron stars.

Further work showed that the spin period of pulsars is increasing, which reflects the fact that pulsars are spinning down over time (see figure 1.1 for a sketch of the pulsar geometry). A magnetic dipole moment model was proposed to explain the spinning down of pulsars, where the rotational energy of pulsars is lost due magnetic dipole radiation (Pacini 1967; Gold 1968; Ostriker and Gunn 1969). The observational results from the Crab pulsar agreed well with this model and suggested that the required surface magnetic fields falls in the range $10^{11} - 10^{13}$ G for the first detected pulsars. This range has since significantly broadened, first with the discovery of a class of pulsars having periods of several milliseconds (Backer et al. 1982), believed to have been spun-up by accretion torques of a binary companion (Alpar et al. 1982), and much lower surface magnetic fields in the range $10^{8} - 10^{10}$ G. Recent surveys have also discovered pulsars with very high period derivatives (e.g. Morris et al. 2002) that imply surface fields up to around $10^{14}$ G. The rotating magnetic dipole model accounts for the spin-down of the pulsar where the rate of rotational energy loss and magnetic field are given by:

$$\dot{E} = \frac{d}{dt}\left(\frac{1}{2}I\Omega^2\right) = I\Omega\dot{\Omega}$$
$$= \frac{2}{3c^3}|m|^2 \Omega^4 \sin^2\alpha \quad (1.1)$$

$$B = \sqrt{\frac{3c^3}{8\pi^2} \frac{I}{R^6 \sin^2\alpha} P\dot{P}} = 3.2 \times 10^{19} \sqrt{P\dot{P}} \text{ Gauss} \quad (1.2)$$

Where $I$ is the neutron star moment of inertia, $\Omega = 2\pi/P$ is angular spin frequency, $P$ is the spin period and $R$ is the radius of the neutron star.

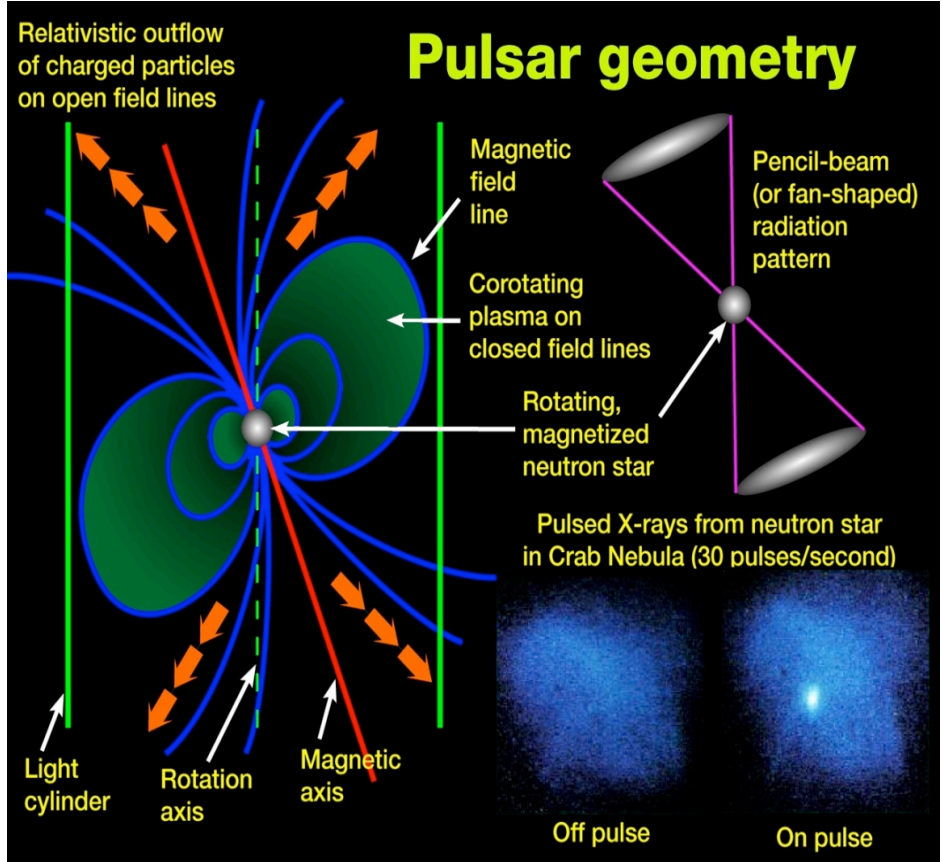

Figure 1.1 Illustrative sketch of the Crab pulsar geometry. It includes an intense magnetic field whose rotation (with the star) controls and drives the activities of plasma for billions of miles around the Crab pulsar (Credit: NASA/Marshall Space Flight Center).

The periods and period derivatives of the various types of isolated neutron stars are shown in the $P$ - $\dot{P}$ diagram of figure 1.2. Assuming that the spin-down rotation is due to magnetic dipole radiation, two quantities can be defined from the measured $P$ and $\dot{P}$ for each neutron star (1) Characteristic age: $\left(P/2\dot{P}\right)$ from $\dot{\Omega} \propto -\Omega^3$ (where $\Omega = 2\pi/P$), the age of the pulsar is found to be $T = \left(P/2\dot{P}\right)\left[1 - \left(P/P_i\right)^2\right]$, where $P_i$ is the pulsar's initial spin period. (2) Surface dipole magnetic field:

$$B_s = \left(\frac{3Ic^3 P\dot{P}}{2\pi^2 R^6}\right)^{1/2} \simeq 2 \times 10^{12} \sqrt{P\dot{P}_{15}} \text{ G} \tag{1.3}$$

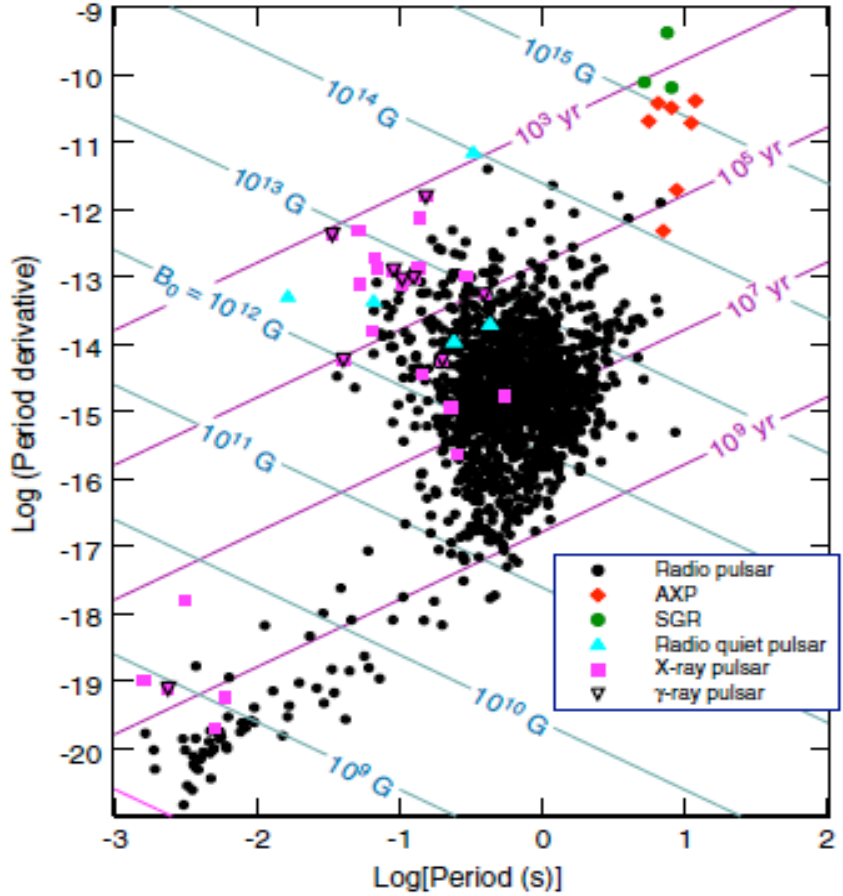

Figure 1.2 Plot of the spin period versus spin period derivative for the presently known rotation-powered pulsars and magnetars. Lines of constant characteristic age, $\left(P/2\dot{P}\right)$, and surface dipole field (see equation 1.3) are superposed (Credit: Harding and Lai 2006).

The magnetars, eight AXPs and four SGRs, occupy the upper right-hand corner of the diagram and curiously somewhat overlap with the region occupied by the high-field radio pulsars. However, the two types of objects display very different observational behavior. The high-field radio pulsars have very weak or non-detectable x-ray emission and do not burst (e.g. Kaspi and McLaughlin 2005), while the magnetars have no detectable radio pulsations, with the exception of the recent detection of radio pulsations in the transient AXP, XTE J1810-197 (Camilo *et al* 2006). At the present time, the intrinsic property that actually distinguishes magnetars from radio pulsars remains to be vague and ambiguous.

## 1.1.3  Birth and Formation of neutron stars

Neutron stars are final products of stellar evolution. It is widely accepted that they are born in supernova explosions after their presupernova progenitors (giant or supergiant stars) exhaust nuclear fuel in their cores. The cores undergo gravitational collapse into neutron stars (or black holes), while outer presupernova layers are blown away by an expanding shock wave, producing supernova remnants. The gravitational collapse of the core of a massive star during a Type II supernova event forms a neutron star. Such stars are composed almost entirely of neutrons. Neutron stars are very hot and are supported against further collapse because of the Pauli exclusion principle. This principle states that no two neutrons (or any other fermions particle) can occupy the same quantum state simultaneously.

The whole event is usually referred to as a core-collapse type II supernova explosion (see, e.g., Imshennik & Nadyozhin 1988; Arnett, 1996, and references therein). The neutron star – supernova connection was suggested by Baade and Zwicky in 1934 as described in section 1.2. The explosion, which occurs in the presupernova core, triggers a shock wave propagating outward (after bouncing off the dense core). It takes several hours for the shock to travel through extended presupernova outer layers. At this stage the presupernova, observed from outside, looks just as usual, as if nothing happened in its interior. After the shock reaches the surface, it produces a splash of radiation in all bands of electromagnetic spectrum to be observed as a supernova event. In addition, the core collapse itself should be accompanied by a powerful outburst of neutrino emission and, possibly, of gravitational radiation. These events could be detectable by neutrino and gravitational observatories prior to the electromagnetic outburst.

Supernova explosions are accompanied by an enormous energy release, a few times $10^{53}$ erg in total (of the order of the gravitational energy of a neutron star, Eq. (1.4)). It is expected that the energy be mostly released in the form of neutrinos. About 1% of the total energy transforms into the kinetic energy of the explosion ejecta, and only a minor part (~ $10^{49}$ erg) into electromagnetic radiation; a smaller part can be emitted in the form of gravitational waves.

## 1.1.4 Properties and Structure of Neutron Stars

Neutron stars are compact stars that contain matter of super-nuclear density in their interiors (presumably with a large fraction of neutrons). They have typical masses $M \sim 1.4\, M_\odot$ and radii $R \sim 10$ km. Thus, their masses are close to the solar mass $M_\odot = 1.989 \times 10^{33}$ g, but their radii are $\sim 10^5$ times smaller than the solar radius $R_\odot = 6.96 \times 10^5$ km. Accordingly, neutron stars possess an enormous gravitational energy $E_{\text{grav}}$ and surface gravity $g$,

$$\begin{aligned} E_{grav} &\sim GM^2/R \sim 5 \times 10^{53} \text{ erg} \sim 0.2 Mc^2, \\ g &\sim GM/R^2 \sim 2 \times 10^{14} \text{ cm/s}^2 \end{aligned} \quad (1.4)$$

Where $G$ is the gravitational constant and $c$ is the speed of light. Clearly, neutron stars are very dense. Their mean mass density is

$$\bar{\rho} \simeq \frac{3M}{4\pi R^3} \simeq 7 \times 10^{14} \text{ g/cm}^3 \sim (2-3)\rho_o \quad (1.5)$$

Where $\rho_0 = 2.8 \times 10^{14}$ g cm$^{-3}$ is the so-called normal nuclear density, the mass density of nucleon matter in heavy atomic nuclei. The central density of neutron stars is even larger, reaching (10 −20) $\rho_0$. By all means, neutron stars are the most compact stars known in the Universe.

Depending on star mass and rotational frequency, the matter in the core regions of neutron stars may be compressed to densities that are up to an order of magnitude greater than the density of ordinary atomic nuclei (for a review see Weber, Negreiros and Rosenfield (2007) and references therein). This extreme compression provides a high-pressure environment in which numerous subatomic particle processes are likely to compete with each other. The most spectacular ones stretch from the generation of hyperons and baryon resonances (Σ, Λ, Ξ, Δ), to quark ($u, d, s$) deconfinement, to the formation of boson condensates ($\pi^-$, $K^-$, H-matter) as shown in figure 1.3. In the framework of the strange matter hypothesis, it has also been suggested that 3-flavor strange quark matter–made of absolutely stable u, d, and s quarks–may be more stable than ordinary atomic nuclei. In the latter event, neutron stars should in fact be made of such matter rather than ordinary (confined) hadronic matter.

Another striking implication of the strange matter hypothesis is the possible existence of a new class of white dwarfs- like strange stars (strange dwarfs). The quark matter in neutron stars, strange stars, or strange dwarfs ought to be in a color superconducting state. This fascinating possibility has renewed tremendous interest in the physics of neutron stars and the physics and astrophysics of strange quark matter.

According to current theories, a neutron star can be subdivided into the atmosphere and four main internal regions: the outer crust, the inner crust, the outer core, and the inner core as shown in Figure 1.3 (see Haensel, Potekhin and Yakovlev (2007) for a full account). The atmosphere is a thin plasma layer, where the spectrum of thermal electromagnetic neutron star radiation is formed. The spectrum, beaming and polarization of emerging radiation can be determined theoretically by solving the radiation transfer problem in atmospheric layers. This radiation contains valuable information on the parameters of the surface layer (on the effective surface temperature, surface gravity, chemical composition, strength and geometry of the surface magnetic field) and on the masses and radii of neutron stars. For temperatures less than ~ 0.1 MeV, the neutron fluid in the crust probably forms a $^1S_0$ superfluid (Baym & Pethick 1975; (Baym, Pethick, & Pines 1969; Østgaard 1971). Such a superfluid would alter the specific heat and the neutrino emissivities of the crust, thereby affecting how neutron stars cool. The superfluid would also form a reservoir of angular momentum that, being loosely coupled to the crust, could cause pulsar glitch phenomena (Anderson & Itoh 1975).

The core constitutes up to 99% of the mass of the star. The outer core consists of a soup of nucleons, electrons, and muons. The neutrons could form a $^3P_2$ superfluid and the protons a $^1S_0$ superconductor within the outer core. In the inner core, exotic particles such as strangeness-bearing hyperons and/or Bose condensates (pions or kaons) may become abundant. It is possible that a transition to a mixed phase of hadronic and deconfined quark matter develops (Glendenning, 1992), even if strange quark matter is not the ultimate ground state of matter. Delineating the phase structure of dense cold quark matter (Alford 2001), has yielded novel states of matter, including color-superconducting phases with (Bedaque & Schäfer, 2002) and without condensed mesons (Alford 2001).

The atmosphere thickness varies from some ten centimeters in a hot neutron star (with the effective surface temperature $T_s \sim 3 \times 10^6$ K) to a few millimeters in a cold one ($T_s \sim 3 \times 10^5$ K). Very cold or ultra-magnetized neutron stars may have a solid or liquid surface. Neutron star atmospheres have been studied theoretically in a number of papers (see, e.g., Zavlin and Pavlov 2002 and references therein). Current atmosphere models, especially for neutron stars with surface temperatures $T_s \leq 10^6$ K and strong magnetic fields $B \geq 10^{11}$ G, are far from being complete. The most serious problems consist in calculating the EOS, ionization equilibrium, and spectral opacity of the atmospheric plasma. If the radiation flux is too strong, the radiative force exceeds the gravitational one and makes the atmosphere unstable with respect to a plasma outflow. In a hot non-magnetized atmosphere, Thomson scattering produces a radiative force, this happens whenever the stellar luminosity $L$ exceeds the Eddington limit

$$L_{Edd} = 4\pi c G M m_p / \sigma_T \approx 1.3 \times 10^{38} \left( M/M \right) \text{ erg s}^{-1} \tag{1.5}$$

Where $\sigma_T$ is the Thomson scattering cross section and $m_p$ is the proton mass.

The outer crust (the outer envelope) extends from the atmosphere bottom to the layer of the density $\rho = \rho_{ND} \approx 4 \times 10^{11}$ g cm$^{-3}$. Its thickness is some hundred meters and its matter consists of ions $Z$ and electrons $e$. A very thin surface layer (up to few meters in a hot star) contains a non-degenerate electron gas. In deeper layers the electrons constitute a strongly degenerate, almost ideal gas, which becomes ultra-relativistic at $\rho \gg 10^6$ g cm$^{-3}$, where electrons mainly provide the pressure. For $\rho \geq 10^4$ g cm$^{-3}$, atoms are fully ionized by the electron pressure. In the outer atmosphere layers the ions may constitute a Boltzmann gas, but in deeper layers they form a strongly coupled Coulomb system (liquid or solid). A larger fraction of the envelope is usually solidified; hence, the envelope is often called the crust. The electron Fermi energy grows with increasing $\rho$. This induces beta captures in atomic nuclei and enriches the nuclei with neutrons. At the base of the outer crust the neutrons start to drip out from the nuclei producing free neutron gas.

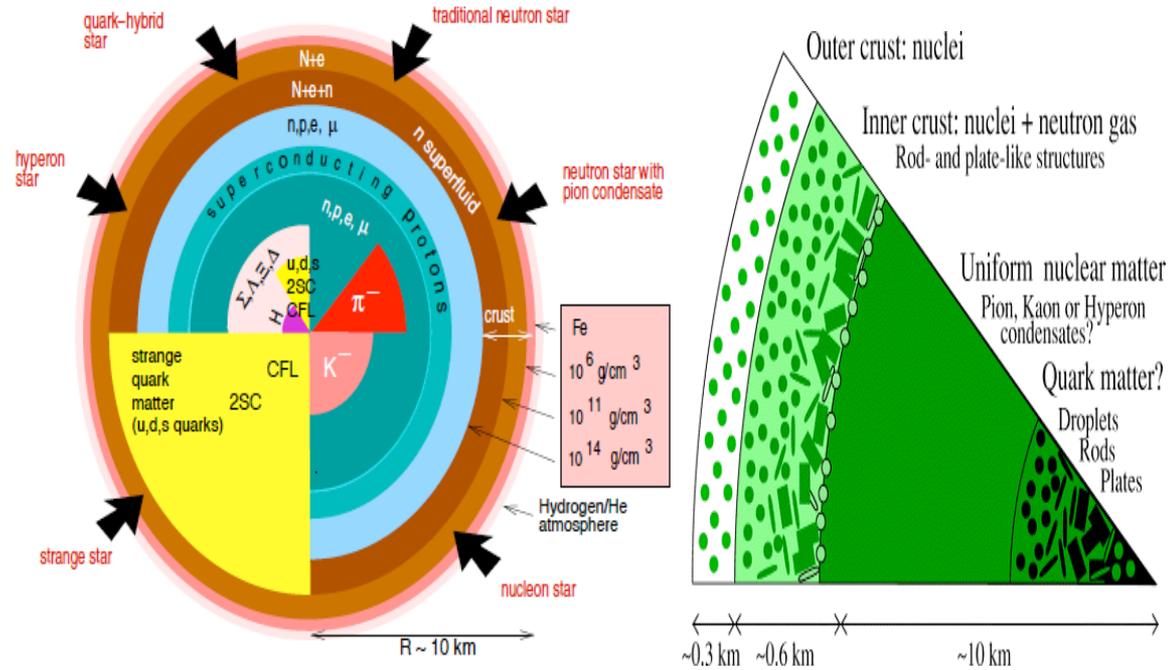

Figure 1.3 Neutron Star structure and its composition as predicted by theory are shown in the left panel (Credit: Weber et al. 2007). Cross-section of a ~ 1.4$M_\odot$ neutron star, where the ~ 1 km thick crust consists of neutron rich nuclei in a lattice and a uniform background of electrons and, in the inner crust, also a neutron gas. (Credit: Heiselberg 2002)

The inner crust (the inner envelope) may be about one kilometer thick. The density $\rho$ in the inner crust varies from $\rho_{ND}$ at the upper boundary to ~ 0.5$\rho_0$ at the base. Here, $\rho_0$ is the saturation nuclear matter density. The matter of the inner crust consists of electrons, free neutrons $n$, and neutron-rich atomic nuclei. The fraction of free neutrons increases with growing $\rho$. The neutronization at $\rho \approx \rho_{ND}$ greatly softens the EOS, but at the crust bottom the repulsive short-range component of the neutron-neutron interaction comes into play and introduces a considerable stiffness. In the bottom layers of the crust, in the density range from $\approx \rho_0/3$ to $\approx \rho_0/2$, the nuclei may become essentially non-spherical and form a "mantle", but this result is model dependent. The nuclei disappear at the crust-core interface. Free neutrons in the inner crust and nucleons confined in the atomic nuclei can be in superfluid state.

The outer core occupies the density range 0.5 $\rho_0 \leq \rho \leq$ 2 $\rho_0$ and is several kilometers thick. Its matter consists of neutrons with several percent a mixture of protons $p$, electrons $e$, and possibly muons $\mu$ (the so called *npeμ* composition). The state of this matter is determined by the conditions of electric neutrality and beta equilibrium, supplemented by a microscopic model of many-body nucleon interaction. The beta equilibrium implies the equilibrium with respect to the beta (muon) decay of neutrons and inverse processes. All *npeμ*-plasma components are strongly degenerate. The electrons and muons form almost ideal Fermi gases. The neutrons and protons, which interact via nuclear forces, constitute a strongly interacting Fermi liquid and can be in superfluid state.

The inner core, where $\rho \geq 2 \rho_0$, occupies the central regions of massive neutron stars (and does not occur in low-mass stars whose outer core extends to the very center). The interior of the neutron star contains a nuclear liquid of mainly neutrons and ~ 10% protons at densities above nuclear matter density n0 increasing towards the center. Here pressures and densities may be sufficiently high that the dense cold strongly interacting matter undergoes phase transitions to, e.g., quark or hyperon matter or pion or kaon condensates appear. Typical sizes of the nuclear and quark matter structures are ~ $10^{-14}$ m. Its radius can reach several kilometers, and its central density can be as high as $(10-15) \rho_0$. The composition and EOS of the neutron star are strongly model dependent. Several hypotheses have been put forward, predicting the appearance of new fermions and/or boson condensates. The main four hypotheses are:

(1) Hyperonization of matter – the appearance of hyperons, first of all $\Sigma^-$ and $\Lambda$ hyperons.

(*2*) Pion condensation – the appearance of a boson condensate of pion-like excitations with a strong renormalization and mixing of nucleon states.

(3) Kaon condensation – the Bose-Einstein condensation of kaon-like excitations, which, like real kaons, possess strangeness.

(4) A phase transition to the quark matter composed of deconfined light *u* and *d* quarks and strange *s* quarks, and a small mixture of electrons, or even no electrons at all.

Nucleon and nucleon/hyperon matter, called respectively nuclear and hypernuclear matter, have been studied experimentally in ordinary nuclei and hypernuclei. Pion and kaon condensations have not been discovered in laboratory so far. Some very tentative signatures of quark deconfinement have been recently detected in the relativistic heavy-ion collisions. The models (2) – (4) are often called exotic models of dense matter. A new phase may appear via a first-order or a second-order phase transition. Its appearance is accompanied by the softening of the EOS. One cannot exclude the existence of mixed phases of dense matter.

Let us mention a special hypothetical class of compact stars, which are called strange stars (see Chapter 4). They could exist only if the absolute ground state of hadronic matter is a self-bound quark matter. Strange stars entirely (or nearly entirely) consist of strange quark matter. In some models, this matter extends to the very surface; such stars are called bare strange stars. In other models, strange stars have a normal crust extending from the surface not deeper than to the neutron-drip density $\rho_{ND}$.

### 1.1.5 Soft gamma repeaters and Anomalous X-ray pulsars

Soft gamma repeaters (SGR) and anomalous X-ray pulsars (AXP) are two types of isolated neutron stars. They seem to form a larger class of magnetars (see, e.g., Thompson 2002, Kaspi 2004, and references therein). In this thesis, the object we are investigating is an SGR (SGR 1806-20). SGRs are sources of repeating soft gamma ray and X-ray bursts. Typical bursts last for ~ 0.1 s and have energies ~ $10^{41}$ erg. Their bursting activity is highly irregular. Years of quiet states are interlaced with weeks of hundreds of bursts. By 2006 four soft gamma repeaters and two candidates have been discovered. The first discovered object, SGR 0525–66, is in the Large Magellanic Cloud, whereas other ones are in the Galactic plane. The most remarkable events were three gigantic gamma-ray bursts, much stronger than typical bursts. The first one was detected from SGR 0525–66 on March 5, 1979 (Mazets *et al.*, 1979), the second one was detected from SGR 1900+14 on August 27, 1998 (Hurley *et al.*, 1999) and the third from SGR 1806–20 on December 27, 2004 (Hurley *et al.* 2005), which had an energy that exceeded $10^{46}$ erg.

Periodic pulsations with large periods, from 5 to 8 s, have been detected in X-rays from the three sources. Two of them show pulsations in quiescent states that have enabled one to measure the spin period derivative $\dot{P}$. In particular, one has got $P = 5.2$ s and $\dot{P} = 6.1 \times 10^{-11}$ s s$^{-1}$ for SGR 1900+14, which gives the characteristic age to be $t = (P/2\dot{P}) \sim 1.3 \times 10^3$ years. Using the values of $P$ and $\dot{P}$, we immediately obtain an enormous surface magnetic field $B_s \sim 5.7 \times 10^{14}$ G. There are other arguments that soft gamma repeaters are young, slowly rotating and rapidly spinning down neutron stars with ultra-strong magnetic fields $B \sim 10^{14}$–$10^{15}$ G.

AXPs are sources of pulsed X-ray emission. The pulsation periods range from 6 to 12 s, and the X-ray luminosities range from $\sim 10^{33}$ to $\sim 10^{35}$ erg s$^{-1}$. These pulsars differ from the classical X-ray pulsars in X-ray binaries by the absence of any evidence that they enter binary systems. By 2005 five AXPs were discovered, together with several candidates. Some of them have been detected in optical. In most of the cases pulsar timing has been performed and the values of $\dot{P}$ have been measured. The estimated characteristic ages are slightly higher than for soft gamma repeaters, but the characteristic magnetic fields are of the same order of magnitude. For instance, for 1E 1048.1–5937 one has $P = 6.4$ s, $\dot{P} = 3.3 \times 10^{-11}$ s s$^{-1}$, $t = (P/2\dot{P}) \sim 3.1 \times 10^3$ years, and $B_s \sim 5.7 \times 10^{14}$ G.

Therefore, AXPs have much in common with soft gamma repeaters. A solid piece of evidence that these sources are related was provided by the discovery of bursting activity of AXPs (in particular, two bursts, separated by 16 days, from 1E 1048.1–5937, Gavriil et al. 2002; and over 80 bursts detected in June 2002 from 1E 2259+586, Kaspi et al. 2003). It is currently assumed that soft gamma repeaters and AXPs belong to the same class of neutron stars, which are called magnetically powered pulsars or magnetars – neutron stars with ultra strong magnetic fields. The magnetar hypothesis was proposed on theoretical grounds, by Duncan & Thompson (1992). SGRs are thought to be younger and transform into a AXPs in the course of their evolution. The sources of both types can be powered by huge magnetic fields located in neutron star interiors. Bursts are thought to be associated with episodic releases of stresses caused by the evolution of magnetic fields in neutron star crusts. The super strong magnetic field is estimated to decay in ~ $10^4$ years hampering the activity of these sources when they become older.

## 1.2 RXTE Space Mission

The Rossi X-ray Timing Explorer (RXTE), named after astronomer Bruno Rossi, probes the physics of cosmic X-ray sources by making sensitive measurements of their variability over time scales ranging from milliseconds to years. How these sources behave over time is a source of important information about processes and structures in white-dwarf stars, X-ray binaries, neutron stars, pulsars, and black holes (see RXTE website on NASA Goddard Space Flight Center).

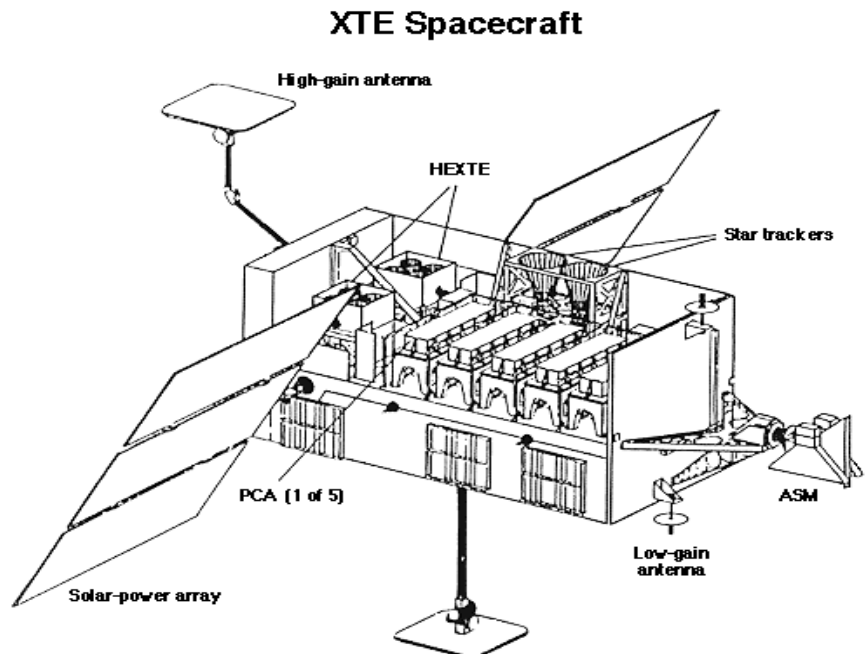

Figure 1.4 Schematic of RXTE space mission. RXTE carries three instruments, PCA, HEXTE and ASM.

The 3000+ kg RXTE satellite was launched on Dec. 30, 1995 atop a Delta II rocket into low-earth orbit (about 600 km and 23 deg inclination). Data links are established through NASA TDRSS satellites, beamed to ground stations, then to Goddard Space Flight Center. The RXTE is maneuverable (6 deg/minute), so that it can be made to point to a chosen source rapidly. This flexibility allows the instrument to respond to short-lived or new phenomena as they are discovered. The instrument cannot point at less than 30 degree toward the sun (due to the Sun's X-ray "brightness".) With instruments sensitive to a wide range of X-ray energies (from 2-200 keV), RXTE is designed for studying known sources, detecting transient events, X-ray bursts, and periodic fluctuations in X-ray emissions (see figure 1.4 for RXTE schematic drawing). The RXTE space mission contains three instruments, The Proportional Counter Array (PCA), The High-Energy X-ray Timing Experiment (HEXTE), and The All Sky Monitor (ASM).

The objectives of RXTE are to investigate:

- Periodic, transient, and burst phenomena in the X-ray emission from a wide variety of objects.

- The characteristics of X-ray binaries, including the masses of the stars, their orbital properties, and the exchange of matter between them.

- The inner structure of neutron stars, and properties of their magnetic fields.

- The behavior of matter just before it falls into a black hole.

- Effects of general relativity that can be seen only near a black hole.

- Properties and effects of super massive black holes in the centers of active galaxies.

- The mechanisms, which cause the emission of X-rays in all these objects.

The RXTE PCA instrument has five xenon gas proportional counter detectors (the X-rays interact with the electrons in the xenon gas) that are sensitive to X-rays with energies from 2-60 keV. The PCA has a large collecting area (6250 cm$^2$). The PCA's pointing area overlaps that of the HEXTE instrument, increasing the collecting area by another 1600 cm$^2$. Events detected by the PCA will be processed on board by the Experiment Data System (EDS) before insertion into the telemetry stream. The High Energy X-ray Timing Experiment (HEXTE) extends the X-ray sensitivity of RXTE up to 200 keV, so that with the PCA, the two together form an excellent high resolution, sensitive X-ray detector. The HEXTE consists of two clusters each containing four phoswich scintillation detectors. Each cluster can ``rock'' (beam switch) along mutually orthogonal directions to provide background measurements 1.5 or 3.0 degrees away from the source every 16 to 128 s. Automatic gain control is provided by using a 241Am radioactive source mounted in each detector's field of view. Events detected by HEXTE will be processed on board by its own data system before insertion into the telemetry stream at an average data rate of 5 kbit/s. Data products include event mode, binned spectra and light curves, and a burst-triggered event buffer. The All Sky Monitor (ASM) rotates in such a way as to scan most of the sky every 1.5 hours, at 2-10 keV, monitoring the long-term behavior of a number of the brightest X-ray sources, and giving observers an opportunity to spot any new phenomenon quickly. The ASM consists of three wide-angle shadow cameras equipped with proportional counters with a total collecting area of 90 cm$^2$. The 3 instruments main properties are summarized in Table (1.1).

Table 1.1 Summary of the main properties for each of the 3 RXTE instruments.

| Instrument Property | PCA | HEXTE | ASM |
|---|---|---|---|
| Energy Range | 2 – 60 keV | 15 – 250 keV | 2 – 10 keV |
| Energy Resolution | < 18% at 6 keV | 15% at 60 keV | 80% of the sky every 90 min |
| Time Resolution | 1 μs | 8 μs | |
| Spatial Resolution | 1 degree FWHM | 1 degree FWHM | 3' x 15' |
| Detectors | 5 proportional counters | 2 clusters of 4 NaI/CsI scintillation counters | Xenon proportional counter |
| Collecting Area | 6500 cm$^2$ | 2 × 800 cm$^2$ | 90 cm$^2$ |
| Sensitivity | 0.1 mCrab[*] | 1 Crab | 20 mCrab |
| Background | 90 mCrab | 50 counts/s | |

[*] mCrab is One thousandth of the intensity of the Crab nebula. This unit is used when comparing observations from different X-ray detectors on different instruments. On RXTE for example 1 mCrab = 2.786 counts/s/pcu (PCA), or 0.250 counts/s per cluster (HEXTE) or 0.075 counts/s per camera (ASM).

Information on events occurring in the PCA detectors is handled by the Experiment Data System (EDS), a microprocessor-driven on-board data system. The EDS is capable of processing up to 500,000 counts per second, and can time the arrival of individual X-rays to about 1 microsecond. The data can be collected in a number of different data modes simultaneously. This facilitates collection and analysis of the data appropriate to different sources. (Data from HEXTE is handled similarly by a separate data system). Data from different EDS and HEXTE data modes and are placed in separate files. From these files, an astronomer extracts light curves and energy spectra. The astrophysicist then applies knowledge about the RXTE instruments to convert measured quantities into physical quantities (e.g. convert channels of a pulse height spectrum into energy measured in kilo-electron Volts (keV)).

Astrophysicists use light curves to analyze short-term or long-term changes in intensity of a source. Further analysis of the light curves may reveal the presence of periodicities or the make-up of complex variations. Such studies may yield the size of the object or of the orbit the object may be in around another body. The energy spectra are analyzed by fitting the data with models of possible energy sources. Statistics determine how well each model fits the data and may be used to eliminate inappropriate models. Successful analysis may determine, for example, the amount of energy emitted by the object and the source of the energy emission. Emission or absorption lines found in the spectrum further measure the composition of material in or surrounding the object.

# Chapter 2

# Physics of Neutron Stars

## 2.1 Introduction

In this chapter, we highlight the underlying physics and main principles governing the peculiar yet intriguing nature of highly magnetized neutron stars and their implications on the neutron star properties. In 1992, it was proposed that Soft gamma repeaters (SGRs) are magnetically powered neutron stars, or magnetars (Duncan and Thompson 1992). Although this claim was first received with some skepticism, subsequent observational studies further endorsed this hypothesis. It is now widely believed that the emission from SGRs is dominantly powered by the magnetic field decay. Kouveliotou et al. (1998) first gave an estimate of the SGR magnetism, the magnetic field strength was estimated to be of the order $8 \times 10^{14}$ Gauss. Magnetars are the most strongly magnetized objects yet known in the universe ($\sim 10^{14} - 10^{15}$ Gauss), their magnetic fields are several orders of magnitude higher than any magnetic field experienced in daily life or the magnetic field strength obtained in the lab (typically $\sim 10^4 - 10^5$ Gauss). At extremely high magnetic fields, several exciting physical effects come into the picture thus making magnetars a hot spot for the investigation and study of such exotic phenomena (see for example the review articles, Duncan 2000; Haensel, Potekhin and Yakovlev (2007); Harding and Lai 2006; Mereghetti

2008; Woods and Thompson 2005)†. [1]

## 2.2 Physics in Ultra-strong Magnetic Fields

Neutron stars represent the domain of extreme physics and states of matter. The strongest curved space-time in the universe, densest non-singular matter and the strongest magnetic fields in the universe are found in neutron stars. The interior of neutron stars has been a source of great mystery and speculation for scientists. The pressure and density inside a neutron star core is thought to be so great that it could harbor exotic particles not made apparent since the moment of the Big Bang. One possibility is that the stars' interiors are home to unbound versions of the building blocks of protons and neutrons, called quarks. Even the most powerful particle accelerators on Earth can't muster up the energies needed to reveal free quarks. This exciting domain of physics offers an incredible opportunity to shed further insight about fundamental physics as well as Astrophysics.

### 2.2.1 Electrons in strong magnetic fields

The problem of quantum mechanics of a charged particle in the presence of a magnetic field is treated in a number of texts (see for example; Landau & Lifshitz (1977), Landau & Lifshitz (1999) and Meszaros (1992)). In this section we summarize the basic results needed for later discussions.

For a non-relativistic particle of charge $e_i$ and mass $m_i$ moving in a uniform magnetic field $B$ oriented along the $z$-axis direction. Classically, the particle will gyrate in a circular orbit with radius and angular frequency given by:

---

† In this chapter I have drawn mostly on the review articles by Duncan (2000), Haensel, Potekhin and Yakovlev (2007) and Harding & Lai (2006) and references therein.

$$\rho = \frac{m_i v_\perp}{|e_i| B}, \qquad \omega_c = \frac{|e_i| B}{m_i} \qquad (2.1)$$

where $v_\perp$ is the particle velocity in the direction perpendicular to the magnetic field. Non-relativistic quantum mechanics dictates that the energy is quantized in Landau Levels

$$E_\perp = \frac{1}{2} m_i v_\perp^2 = \frac{1}{2 m_i} \Pi_\perp^2 = \left(n + \frac{1}{2}\right) \hbar \omega_c, \quad n = 0, 1, 2, \ldots \qquad (2.2)$$

where $\Pi = p - \frac{e_i}{c} A = m_i v$ is the mechanical momentum, $p = -i\hbar \nabla$ is the canoncial momentum and $A$ is the magnetic vector potential.

For an electron $(m_i \to m_e$ and $e_i \to -e)$, the basic quantum energy is the cyclotron energy given by

$$E_{cyclotron} = \hbar \omega_c = \hbar \frac{eB}{m_e c} = 11.577 B_{12} \text{ keV} \qquad (2.3)$$

where $B_{12}$ is the magnetic field in units of $10^{12}$ G. For extremely strong magnetic fields, where $\hbar \omega_c \geq m_e c^2$ is satisfied or

$$B \geq B_Q = m_e^2 c^3 / \hbar e = 4.4 \times 10^{13} \text{ G} \qquad (2.4)$$

The transverse motion of the electron becomes relativistic and the transverse momentum is again quantized according to equation (2.2) i.e. $\Pi_\perp^2 = (2n+1)\hbar e B/c$.

The quantum electrodynamic field strength $B_Q \equiv m_e^2 c^3 / \hbar e = 4.4 \times 10^{13}$ Gauss is very critical in defining a threshold value for the behavior of electron, atoms and molecules. Where $m_e$ is the mass of the electron, $c$ is the speed of light, $\hbar$ is the reduced Planck's constant and $e$ is the electron charge. In magnetic fields, where $B > B_Q$ the electron will gyrate relativistically with quantum gyration radius $r_{gyr} \sim \lambda_e (B/B_Q)^{-1/2}$, where $\lambda_e \equiv \hbar/m_e c$ is the electron Compton wavelength. From quantum mechanics we know that Heisenberg's uncertainty principle quantum mechanics requires that the condition $r.p \sim \hbar$ be satisfied.

Therefore we obtain a canonical momentum $p \sim (\hbar/r_{gyr}) \sim m_e c (B/B_Q)^{1/2}$ causing the electron to gyrate relativistically, where the electron's Landau excitation energy exceeds its rest energy. This can be seen from the solution of Dirac's equation for an electron in a homogenous magnetic field, where the Dirac spinors are proportional to Hermite polynomials and the Landau excitation energy levels are given by:

$$E_n = \left[ c^2 p_z^2 + m_e^2 c^4 \left( 1 + 2n \frac{B}{B_Q} \right) \right]^{1/2} \quad (2.5)$$

The Landau excited energy levels are two-fold degenerate in *s* (spin). The first Landau-level excitation energy is very large; therefore electrons almost always remain in the ground state for processes thought to occur near the surfaces of ultra magnetized neutron stars.

$$\Delta E = E_1 - E_o \approx m_e c^2 \sqrt{2B/B_Q} \quad \text{for } B \gg B_Q \quad (2.6)$$

J. Schwinger estimated the anomalous magnetic moment of the electron and demonstrated that the Electron self-interactions resolve the degeneracies of the Landau energy levels, and shifts the ground state energy (Schwinger 1948). A first order correction to the electron's effective spin magnetic moment is enhanced by the factor $(1 + \alpha/2\pi)$, where $\alpha$ is the fine-structure constant defined as $\alpha \equiv e^2/\hbar c = 1/137$. This yields a ground-state energy shift

$$E_o = m_e c^2 \left[ 1 - (\alpha/2\pi)(B/B_Q) \right]^{1/2} \quad (2.7)$$

The above formula implies that the ground-state energy of the electron goes to zero at $B = (2\pi/\alpha) B_Q \approx 3.8 \times 10^{16}$ Gauss. For stronger magnetic field strength $B$, the vacuum would become unstable to pair production, with remarkable physical consequences (O'Connel 1968; Chiu and Canuto 1968). However equation (2.7) is valid only in the sub-$B_Q$ regime. According to Jancovivi (1969), the asymptotic fractional enhancement (to first order in $\alpha$) valid for very large $B$, is given by

$$(E_o - m_e)/m_e = (\alpha/4\pi)\left(\left[\ln(2B/B_Q) - \xi - \frac{3}{2}\right]^2 + \beta\right) \quad (B \gg B_Q) \tag{2.8}$$

in the god given natural units where $\hbar = c = 1$. The constant $\xi = 0.577$ is Euler's constant, and $\beta = 3.9$ is a numerical constant from evaluating integrals numerically.

Therefore, we see that at magnetic field strength $B \sim 10^{32}$ Gauss, the electron's ground state energy is doubled $(E_o \sim 2 m_e c^2)$. However, the maximum magnetic fields in neutron stars are several orders of magnitude short of this value. Magnetic fields of the order $B \sim 10^{17}$ G is possible in the case of the efficient conversion of the differential rotation free energy of a rapidly rotating, new born neutron star through a post-collapse dynamo. From equation (2.8) at $B \sim 10^{17}$ G, we get $E_o - m_e \approx 0.03 m_e$ which implies that magnetic self-energy corrections for electron and positrons are negligible for the typical range of magnetic fields found in neutron stars.

### 2.2.2 Atoms, Molecules and Matter in strong magnetic fields

At sufficiently low temperatures, a magnetar's surface will be covered with atoms and molecules. This surface structure can have consequences for the star's quiescent X-ray emissions, because it determines the work function for removing charged particles from the surface, as necessary for maintaining currents in the magnetosphere. Such currents may result from magnetically driven crustal deformations such as twists of circular patches of the crust.

If a bundle of field lines, describing an arch in the magnetosphere, has one footpoint twisted (with the motion driven from below by the evolving field), then a current must flow along the arch to maintain the twisted exterior field, since $\oint B.dl = \mu I$. Surface impacts of the flowing charges create hot spots at the arch's footpoints and ultimately dissipate the exterior magnetic energy of the twist, with implications for SGR and AXP X-ray light curves and their time-variation. Here we focus on the atomic and molecular physics that comes into play, following a paper by Ruderman (1974) and extending the arguments to $B > B_Q$.

The Bohr radius of a hydrogen atom is $r_o = \lambda_e/\alpha$. The quantum gyration radius, $r_{gyr} = \lambda_e (B/B_Q)^{-1/2}$, is smaller than $r_o$ for $B > \alpha^2 B_Q = 2.4 \times 10^9$ G. This is the characteristic field strength at which magnetism radically alters the atomic structure of matter. At $B > \alpha^2 B_Q$, an atomic electron is constrained to gyrate along a cylinder, which lies entirely within the spherical volume that the unmagnetized atom would occupy. Electrostatic attraction binds the electron strongly to the central nucleus. At $B \gg \alpha^2 B_Q$, the cylinder becomes very long and narrow, and atomic binding energies are adequately given by eigenvalues of the one-dimensional Schrodinger equation. A simple, intuitive estimate—which gives a good estimate of the ground state energy despite its lack of rigor—involves idealizing the atom as a line-charge of length $2\ell$. For linear charge density $e/2\ell$, the electrostatic energy is $\varepsilon = -(e^2/\ell) \ln[\ell/r_{gyr}]$. A lower cutoff $r_{gyr}$ is necessary because the charge distribution does not resemble a line when you get within $\sim r_{gyr}$ of the nucleus. It is more like a sphere, contributing an energy $\sim -qe/r_{gyr}$ where $q = er_{gyr}/\ell$; but this contribution can be neglected in the limit $\ell \gg r_{gyr}$ or $B \gg \alpha^2 B_Q$. Thus, the ground state energy, including the energy of non-relativistic motion parallel to B, is $E_o(\ell) = (\hbar^2/2m_e\ell^2) - (e^2/\ell) \ln[\ell/r_{gyr}]$. Minimizing this according to $dE_o/d\ell = 0$, we find $\ell \simeq r_o [\ln(r_o/r_{gyr})]^{-1}$. This shows that the length of the thin cylindrical atom is less than the Bohr diameter, but only by a modest, logarithmic factor. The ground state hydrogen binding energy is then

$$E_o \simeq -(\varepsilon_o/4)\left[\ln\left(B/\alpha^2 B_Q\right)\right]^2 \qquad \text{for } B \gg \alpha^2 B_Q \qquad (2.9)$$

where $E_o = \alpha^2 m_e/2 = 13.6$ eV is one Rydberg. Note that $E \propto [\ln B]^2$ energy scaling are ubiquitous in ultra-magnetized systems. As $B$ increases beyond $B \sim B_Q$, the radius of the atomic cylinder shrinks to less than the Compton wavelength but equation (2.9) remains a reasonably good approximation. This is because the electron's inertia for longitudinal motion (∥ B) stays close to me in the ground-state Landau level even at $B > B_Q$. Equation (2.9) would become invalid if the longitudinal motion became relativistic. But this would require $\ell < \lambda_e$, which occurs only at $B > \alpha^2 \exp(2/\alpha) B_Q \approx 10^{115}$ G. Magnetic fields can never get this strong. We will show in the next section that the vacuum breaks down at smaller $B$. Equation (2.9) implies that the binding energy of hydrogen near the surface of a magnetar with $B \simeq 10\, B_Q$, is $E_o \simeq 0.5$ keV. This is comparable to the surface temperatures of some young magnetar candidates (Heyl and Hernquist 1997).

## 2.3 Magnetic Vacuum breakdown

Magnetars are the strongest magnetized objects yet observed and known to exist in our universe with a magnetic field of the order $10^{14} - 10^{15}$ G. The uniformly magnetized vacuum is stable against spontaneous electron-positron pair production up to a certain magnetic field strength. However, theories suggest that at sufficiently high magnetic fields, the vacuum becomes unstable and eventually must break down.

### 2.3.1 Dirac Magnetic Monopoles

In a visionary paper, Paul Dirac showed that quantum mechanics does not really prelude the existence of isolated magnetic monopoles (Dirac 1931). Dirac investigated the connection between the smallest electric charge $e$ and the smallest magnetic monopole charge $\mu$. Since the smallest charge is known to exist experimentally and to have the value $e$ given by

$$hc/e^2 = 137 = 1/\alpha \tag{2.10}$$

In this study, Dirac considered the non-integrable phases for wave functions and showed how the non–integrable derivatives of the phase of the wave function receive a natural interpretation in terms of the potentials of the electromagnetic field, as the result of which our theory becomes mathematically equivalent to the usual one for the motion of an electron in an electromagnetic field and gives us nothing new. Under the condition that $2\pi \sum n + e/hc . \int H.ds$, for any closed surface it follows that the end points of nodal lines must be the same for all wave functions. These end points are then points of singularity in the electromagnetic field. The total flux of magnetic field crossing a small closed surface surrounding one of these points is

$$4\pi\mu = 2\pi nhc/e \tag{2.11}$$

where *n* is the characteristic of the nodal line that ends there, or the sum of the characteristics of all nodal lines ending there when there is more than one. Thus at the end point there will be a magnetic pole of strength

$$\mu = \frac{1}{2} n\hbar c/e \tag{2.12}$$

The theory thus allows isolated magnetic poles, but the strength of such poles must be quantized, the quantum $\mu_o$ being connected with the electronic charge *e* by

$$\hbar c / e\mu_o = 2 \tag{2.13}$$

This equation is to be compared with (2.9). The theory also requires a quantization of electric charge, since any charged particle moving in the field of a pole of strength $\mu_o$ must have for its charge some integral multiple (positive or negative) of *e*, in order that wave functions describing the motion may exist.

Magnetic monopoles with mass $m_\mu$ and magnetic charge $\mu$ are spontaneously created when the energy they acquire in falling across a monopole Compton wavelength, $E \sim \mu B.(\hbar/m_\mu c)$, exceeds their rest energy $m_\mu c^2$. Therefore the maximum magnetic fields is

$$B_{max} \sim \alpha \left(m_\mu/m_e\right)^2 B_Q \tag{2.14}$$

For Planck-mass monopoles, $m_\mu = 10^{19}$ GeV, this sets a firm upper limit on $B_{max} \sim 10^{55}$ G.

### 2.3.2 Grand unification theories and upper limits

Einstein spent the last few decades of his life trying unsuccessfully to unify the fundamental forces of nature in a somewhat deeper theoretical framework. The period of the 1930's up to the 1970's was the Golden Age for fundamental particle physics, where everything was converging towards a simpler picture that eventually became known as 'The Standard Model' of elementary particle physics, which provided a consistent theory for the description of elementary particles. Although the standard model of elementary particles framework proved successful in unifying three of the fundamental forces that governed the microscopic world (electromagnetic, weak and strong interactions), it had a glaring omission that is the most familiar force of all 'Gravity'. The past 30 years have witnessed an incredible boost in theoretical physics towards reaching a deeper framework that would encompass all the fundamental forces; these theories are generally referred to as Grand unification theories (GUT) in which they attempt to resolve the conflict between Quantum Mechanics and the theory of General Relativity. GUT theories predict that $m_\mu = 10^{16}$ GeV and $B_{max} \sim 10^{49}$ G.

Among the different GUT theories, Superstring/M-Theory is the most appealing candidate for a theory of everything that would explain all of the elementary particles and the fundamental forces as well as reconcile quantum mechanics and general relativity. Superstring theory provides an intermediate estimate for the magnetic monopole mass to be $m_\mu = \alpha_s^{-1/2} = 10^{17} - 10^{18}$ GeV, where $\alpha_s$ is the string tension, this yields a maximum magnetic field strength of the order

$$B_{max} \sim 10^{51} - 10^{53} \text{ G} \tag{2.15}$$

A new radical idea proposed the existence of "large" extra dimensions of space could give have an energy scale for quantum gravity as low as $m_o \sim 1$ TeV (Arkani-Hamed, Dimopoulos and Dvali 1998). It proposes a new framework for solving the hierarchy problem, which does not rely on either supersymmetry or Technicolor. In this framework, the gravitational and gauge interactions become united at the weak scale, which we take as the only fundamental short distance scale in nature. The observed weakness of gravity on distances $\geq 1$ mm is due to the existence of $n \geq 2$ new compact spatial dimensions large compared to the weak scale. The Planck scale $m_p \sim G_N^{-1/2}$ is not a fundamental scale; its enormity is simply a consequence of the large size of the new dimensions. While gravitons (the messenger particle responsible for gravitational forces) can freely propagate in the new dimensions, at sub-weak energies the Standard Model fields must be localized to a 4-dimensional manifold of weak scale "thickness" in the extra dimensions. This picture leads to a number of striking signals for accelerator and laboratory experiments. For the case of $n$s2 new dimensions, planned sub-millimeter measurements of gravity may observe the transition from $1/r^2 \to 1/r^4$ Newtonian gravitation.

The extra dimensions are wrapped up in closed geometries (e.g. circles or loops) of size $L \sim l_p (m_o/m_p)^{-(n+2)/2}$ for n extra dimensions, or $L \sim 1(m_o/1 \text{ TeV})^{-2}$ millimeter for 2 extra dimensions; where $m_p = \sqrt{\hbar c/G} = 2.176 \times 10^{-5} g$ is the Planck mass and $l_p = \sqrt{G\hbar/c^3} = 1.616 \times 10^{-33} cm$ is the Planck length. However, at the present time there's no direct experimental evidence for the existence of large extra dimensions yet we hope we can find its signatures at the Large Hadron Collider (LHC). The above conditions set limiting field strength: $B_{max} \simeq 10^{23} \left( m_\mu / 1 \text{ TeV} \right)^2 \text{ G}$, but the upper limit given by equation (2.14) seems more plausible.

### 2.3.3 Superconducting Cosmic Strings

Superconducting cosmic strings are a plausible consequence of symmetry breaking in grand unified gauge theories. The luminosity in electromagnetic radiation of an oscillating current-carrying loop may substantially exceed the luminosity in gravitational radiation. In the typical case considered, the energy released electromagnetically is $10^{49}$ erg s$^{-1}$, or $10^{66}$ erg in Toto. Several consequences follow from this, the most interesting of which is the possibility that such loops may heat their surroundings, generating large, dense spherical shells of gas. Galaxies forming on these gravitationally unstable shells at moderate redshift will be seen at the present epoch to lie on bubbles having radii in the range 10–20$h^{-1}$ Mpc if the initial ratio of luminosity in electromagnetic waves to that in gravitational waves is $> 10^{-3}$ for mass/length $10^{22}$ g cm$^{-1}$. The required primordial energy density in magnetic fields is $> 3 \times 10^{-9}$ of the radiation energy density, if the charge carriers are bosons or super heavy fermions. Since these shells fill up space, the galaxies will have a distribution similar to that found in a recent survey of the northern sky. When the current saturates, a loop will emit particles copiously, and may be seen as an X-ray source at $z$ ∼10–50. Such loops may also contribute significantly to the hard X-ray and γ-ray backgrounds and to $10^{20}$ eV cosmic rays.

In the previous sections we have seen that a vast range of tremendous magnetic field strength are possible in Nature and GUT theories set an upper limit on the maximum field strength that can exist before driving vacuum unstable, causing it to eventually breakdown. We don't yet know any objects that generate such fields, but some possibilities have been suggested. For example, superconducting cosmic strings—if they exist—could generate fields $\geq 10^{30}$ G in their vicinities (Ostriker, Thompson and Witten 1986). Perhaps future astrophysicists will regard neutron star magnetic fields as mild!

## 2.4  Neutron star magnetic field evolution

Magnetic field is perhaps the single most important quantity that determines the various observational manifestations of neutron stars. Thus it is natural that a large amount of work has been devoted to the study of neutron star magnetic field evolution. Recent reviews include Bhattacharya and Srinivasan (1995), Ruderman (2004) and Reisenegger *et al* (2005).

### 2.4.1  Origin of the neutron star magnetic field

The magnetic fields of neutron stars were most likely already present at birth. The traditional fossil field hypothesis suggests that the magnetic field is inherited from the progenitor, with magnetic flux conserved and field amplified ($B \propto R^{-2}$) during core collapse. In the case of magnetic white dwarfs (with measured fields in the range $\sim 3 \times 10^4 - 10^9$ G), there is strong evidence that the fields are the remnants from a main-sequence phase (Ap/Bp stars, with $B \sim 200$ G – 25 kG) (see e.g. Ferrario and Wichramasinghe (2005)). Neutron stars descend from main-sequence stars with mass $\geq 8\ M_\odot$. Recently, the large-scale magnetic fields (with $B \sim 1$ kG) of O stars were detected (in two stars so far; see Donati *et al* (2006)). It is interesting that the magnetic flux of such O stars $\Phi \sim 10^5 \pi R_\odot$ G (for $R \sim 10\ R_\odot$; of course, not all of this flux threads the inner 1.4 $M_\odot$ core) is of the same order of magnitude as the flux of a $10^{15}$ G neutron star ($R \sim 10^{-5}\ R_\odot$) as well as the fluxes of the most strongly magnetic Ap/Bp stars and white dwarfs. It is also of interest to note that magnetic white dwarfs ($B \geq 1$ MG) tend to be more massive (mean mass $\sim 0.93\ M_\odot$) than their non-magnetic counterparts (mean mass $\sim 0.6\ M_\odot$). Since white dwarfs with $M \geq 0.7\ M_\odot$ and neutron stars form exclusively from the material that belongs to the convective core of a main sequence star (Reisenegger 2001), this suggests that the magnetic field may be generated in the convective core of the main-sequence progenitor. It has also been suggested that magnetized neutron stars could be produced, in principle, by accretion-induced collapse of magnetic white dwarfs (Usov 1992).

Alternatively, it has been argued that magnetic field may be generated by a convective dynamo in the first ~10 s of a proto-neutron star (Thompson and Duncan 1993). Most Astrophysicists supposed that the magnetic field is a relic of the time before the star went supernova. All stars have weak magnetic fields, and those fields can be amplified simply by the act of compression. According to Maxwell's equations of electromagnetism, electromagnetism, as a magnetized object shrinks by a factor of two, its magnetic field strengthens by a factor of four. The core of a massive star collapses by a factor of $10^5$ from its birth through neutron star formation, so its magnetic field should become $10^{10}$ times stronger. If the core magnetic field started with sufficient strength, this compression could explain pulsar magnetism. Unfortunately, the magnetic field deep inside a star cannot be measured, so this simple hypothesis cannot be tested. There are also good reasons to believe that compression is only part of the story.

Within a neutron star, gas can circulate by convection. Warm parcels of ionized gas rise, and cold ones sink. Because ionized gas conducts electricity well, any magnetic field lines threading the gas are dragged with it as it moves. The field can thus be reworked and sometimes amplified. This phenomenon, known as dynamo action, is thought to generate the magnetic fields of stars and planets. A dynamo might operate during each phase of the life of a massive star, as long as the turbulent core is rotating rapidly enough. Moreover, during a brief period after the core of the star turns into a neutron star, convection is especially violent. In principle, the maximum field achievable is either $B \sim (4\pi\rho)^{1/2} v_{con} \sim 4 \times 10^{15}$ G (for convective eddy speed $v_{con} \sim 10^3$ km s$^{-1}$) or $B \sim (4\pi\rho)^{1/2} R\Delta\Omega \sim 2 \times 10^{17}$ G (for differential rotation $\Delta\Omega \sim 2\pi/(1ms)$). How much dipole field can be generated is more uncertain. It could be that a large-scale field of ~$10^{15}$ G is generated if the initial spin period of the neutron star is comparable to the convective turnover time (~ $H_p/v_{con}$ ~ 1ms for pressure scale height of 1 km), whereas in general only a small-scale (~ $H_p$) field is produced, resulting in a mean field of order $10^{12}$–$10^{13}$ G typical of radio pulsars. We note that studies of presupernova evolution of massive stars, including magnetic fields, typically find that a newly formed neutron star has a rotation rate appreciably slower than the breakup rate (Heger *et al* 2005). In either the fossil field or dynamo scenario, the magnetic field of a proto-neutron star is likely 'messy' and unstable. Before the crust forms (around 100 s after collapse), the field will evolve on the Alfven crossing time $t_A \sim (4\pi\rho)^{1/2}R/B$ (~0.1 s for $B \sim 10^{15}$ G) into a stable configuration in which a poloidal field coexists with a toroidal field of the same order of magnitude (see Braithwaite and Spruit (2006), we discuss it in more detail in Chapter 4). Gradual field generation due to a thermomagnetic effect in the crust has been discussed (Urpin and Yakovlev 1980, Blandford *et al* 1983), but it seems unlikely to be capable of producing large-scale fields stronger than $10^{12}$ G.

### 2.4.2 Observations

As discussed earlier, with a few exceptions, our current knowledge of neutron star magnetic fields are largely based on indirect inferences.

(i) For radio pulsars, of which about 1600 are known today (see Manchester (2004)), the period $P$ and period derivative $\dot{P}$ can be measured by timing the arrivals of radio pulses. If we assume that the pulsar spin down is due to magnetic dipole radiation, we obtain an estimate of the neutron star surface field $B$ (see equation (1.3)). For most radio pulsars, the inferred $B$ lies in the range $10^{11}$–$10^{14}$ G. For a smaller population of the older millisecond pulsars, we have $B \sim 10^{8-9}$ G. Such a field reduction is thought to be related to the recycling process that turns a dead pulsar into an active, millisecond pulsar.

(ii) For anomalous x-ray pulsars and soft gamma repeaters (Woods and Thompson 2005), superstrong magnetic fields ($B \sim 10^{14} - 10^{15}$ G) are again inferred from the measured $P$ (in the range 5–12 s) and $\dot{P}$ (based on x-ray timing study) and the assumption that the spin-down is due to the usual magnetic dipole radiation or Alfven wave emission. In addition, strong theoretical arguments have been put forward (Thompson and Duncan 1995, 1996, 2001) which suggest that a superstrong magnetic field is needed to explain various observed properties of AXPs and SGRs (e.g. energetics of SGR bursts/flares, including the spectacular giant flares from three SGRs: the March 5 1979 flare of SGR 0525-66 with energy $E \geq 6 \times 10^{44}$ erg, the August 27 1998 flare from SGR 1900+14 with $E \geq 2 \times 10^{44}$ erg and the December 27 2004 flare from SGR 1806-20 with $E \sim 4 \times 10^{46}$ erg; quiescent luminosity $L_x \sim 10^{34\text{-}36}$ erg s$^{-1}$, much larger than the spindown luminosity $I\Omega\dot{\Omega}$). Tentative detections of spectral features during SGR/AXP bursts have been reported in several systems (e.g. Gavriil *et al* (2002), Ibrahim *et al* (2003), Rea *et al* (2003)), which, when interpreted as proton cyclotron lines, imply $B \sim 10^{15}$ G.

### 2.4.3 Physics of Magnetic fields evolution

The magnetic field in a neutron star evolves through a series of quasi-equilibrium states, punctuated by the release of elastic stress in the crust and hydrodynamic motions in its fluid core. The physics of the quasi-equilibrium field evolution was discussed by Goldreich and Reisenegger (1992) (see also Reisenegger *et al* (2005)). The bulk region of a neutron star comprises a liquid core of mostly neutrons with a small fraction ($Y_e \sim$ a few %) of protons and electrons. The medium is stably stratified due to the $Y_e$ gradient, thus the magnetic field cannot force the bulk neutron fluid to move (e.g. due to buoyancy) unless $B^2/(8\pi) \geq Y_e P$ (or $B \geq 10^{17}$ G) or the fluid can change its composition as it rises (which takes place on a timescale longer than the neutrino cooling time for $B \leq 10^{17}$ G). The magnetic field evolves according to

$$\frac{\partial B}{\partial t} = -\nabla \times \left( \frac{c^2}{4\pi\sigma} \nabla \times B \right) + \nabla \times \left( -\frac{j}{n_e e} \times B \right) + \nabla \times (v_a \times B) \qquad (2.16)$$

where the three terms on the right-hand side represent three different effects.

(i) The first term represents Ohmic diffusion of the magnetic field ($\sigma$ is the zero-field conductivity defined by $j = \sigma E$), with timescale

$$t_{Ohmic} \sim \frac{4\pi\sigma L^2}{c^2} \qquad (2.17)$$

For fields that vary on a lengthscale $L$. In the core, the conductivity is large since all the particles are degenerate, leading to $t_{Ohmic} \sim 2 \times 10^{11} (L_{km}/T_8)^2 (\rho/\rho_{nuc})^3$ yr (Baym *et al* 1969), where $\rho_{nuc} = 2.8$ $\times\ 10^{14}$ g cm$^{-3}$, $L_{km} = L/(1\ \text{km})$. Of course, currents that are confined to the crust would have a much shorter decay time (e.g. Sang and Chanmugam (1987), Cumming *et al* (2004)).

(ii) The second term describes advection of the field by Hall drift: the magnetic field is carried by the electron fluid, which drifts with respect to the ions with velocity $v_e = -j/(n_e e)$. This term is non-dissipative but can change the field structure on timescale

$$t_{Hall} \sim \frac{4\pi n_e e L^2}{cB} \sim 5\times 10^8 \frac{L_{km}^2}{B_{12}}\left(\frac{\rho}{\rho_{nuc}}\right)\ \text{years} \tag{2.18}$$

Goldreich and Reisenegger (1992) suggested that the nonlinear Hall term may give rise to a turbulent cascade to small scale, thus enhancing the Ohmic dissipation rate of the field. This 'Hall cascade' has been confirmed by recent simulations of electron MHD turbulence (also known as whistler turbulence; see Cho and Lazarian (2004)). Several studies have also demonstrated magnetic energy transfer between different scales (e.g. Urpin and Shalybkov (1999). A specific example of the Hall-drift enhanced Ohmic dissipation can be seen by considering a purely toroidal field (see Reisenegger *et al* (2005)): in this case, the evolution equation for the field reduces to the Burgers equation (see also Vainshtein *et al* 2000), where the field tends to develop a cusp (and the associated current sheet) and fast Ohmic decay.

Recently, Rheinhardt and Geppert (2002) and Rheinhardt *et al* (2004) discussed a 'Hall drift instability' that can lead to nonlocal transfer of magnetic energy to small scales, leading to enhanced crustal field dissipation.

(iii) The third term describes ambipolar diffusion, which involves a drift of the combined magnetic field and the bulk electron–proton fluid relative to the neutrons. The drift velocity $v_a$ is determined by force balance $m_p v_a/\tau_{pn} = f_B - \nabla(\Delta\mu)$, where $\tau_{pn}$ is the proton–neutron collision time, $f_B = j \times B/(cn_p)$ is the magnetic force per proton–electron pair, and $\nabla(\Delta\mu)$ (with $\Delta\mu = \mu_p + \mu_e - \mu_n$) is the net pressure force due to imbalance of the $\beta$-equilibrium (Goldreich and Reisenegger 1992). There are two modes of ambipolar diffusion: the solenoidal mode ($\nabla \cdot v_a = 0$) is non-compressive and does not perturb β-equilibrium, thus $v_a = \tau_{pn} f_B/m_p$, and the associated diffusion time $t^s_{Amb} \sim L/v_a$ is

$$t^s_{Amb} \sim \frac{4\pi n_p m_p L^2}{\tau_{pn} B^2} \sim 3 \times 10^9 \frac{T_8^2 L_{km}^2}{B_{12}^2} \text{ year} \quad (2.19)$$

where we have used $n_p \sim 0.05\, \rho/m_p$ and $\tau_{pn} \sim 2 \times 10^{-17} T_8^{-2} (\rho/\rho_{nuc})^{1/3}$ s. The irrotational mode ($\nabla \times v_a = 0$) is compressive and is impeded by chemical potential gradients $\nabla(\Delta\mu)$ and is possible only if weak interaction re-establishes β-equilibrium during the drift. This gives a timescale $t^{ir}_{Amb} \sim t_{cool}/B^2_{17} \sim 5 \times 10^{15} T_8^{-6} B_{12}^{-2}$ yr (where $t_{cool}$ is the neutrino cooling time and modified URCA rate has been used). Like Ohmic decay, ambipolar diffusion is dissipative, which leads to heating of the core and deep crust of a magnetar; this may be the power source for the persistent x-ray emission of magnetars (Thompson and Duncan (1996), Heyl and Kulkarni (1998), Arras et al (2004)).

In addition to the 'steady' field evolution discussed above, crust fracture can also lead to sudden change of the magnetic field (Thompson and Duncan 1996). The crust has a finite shear modulus $\mu$, and when the yield strain $\theta_{max}$ is exceeded the lattice will fracture. The characteristic yield field strength is of order $B_{char} \sim (4\pi\theta_{max}\mu)^{1/2} = 2 \times 10^{14} (\theta_{max}/10^{-3})^{1/2}$ G. For example, Hall drift of the magnetic field causes the stresses in the crust to build up, and irregularities in the field can be damped quickly by crustal yielding—this may be responsible for magnetar bursts (Thompson and Duncan (1995)).

# Chapter 3

# In Search for QPOs in Magnetars

## 3.1 Introduction

In this chapter, we describe in detail our analysis methodology and our main results in our attempt to detect quasi-periodic oscillations in the recurrent burst emission from SGR 1806-20 during October and November 1996. We present evidence for the existence of Quasi-Periodic Oscillations (QPOs) in the recurrent outburst activity from the soft gamma repeater SGR 1806-20 using Rossi X-ray Timing Explorer (RXTE) observations. By searching a large sample of bursts for timing signals at the frequencies of the QPOs discovered in the 2004 December 27 giant are from the source, we find three QPOs at 84, 103, and 648 Hz in three different bursts. The first two QPOs lie within 8.85% and 11.83%, respectively, from the 92 Hz QPO detected in the giant flare. The third QPO lie within 3.75% from the 625 Hz QPO also detected in the same flare. The detected QPOs are found in bursts with different durations, morphologies, and brightness, and are vindicated by Monte Carlo simulations. The fact that we can find evidence for QPOs in the recurrent bursts at frequencies relatively close to those found in the giant are is intriguing and can offer insight about the origin of the oscillations.

## 3.2 Giant flare emission from Magnetars

In a prophetic paper, Duncan and Thompson (1992) argued that magnetars are a special class of neutron stars with ultra-strong magnetic fields of the order $10^{14} - 10^{15}$ Gauss. The suggested magnetar model proved very successful in explaining a wide range of phenomenology of Soft gamma-ray repeaters (SGRs) and Anomalous X-ray Pulsars (AXPs). The main distinction is that SGRs are the only known source of giant flares, which are the brightest cosmic events originating outside the solar system, in terms of the flux received at Earth. SGRs are also characterized by the repetitive emission of bright bursts of low-energy gamma rays with short durations (~ 0.1 s) and peak luminosities reaching up to $10^{41}$ erg s$^{-1}$, well above the classical Eddington luminosity $L_{edd} \approx 2 \times 10^{38}$ erg s$^{-1}$ for a 1.4 $M_\odot$ neutron star. They are also persistent X-ray pulsars with slow spin periods ($P$ ~ 2.6 – 9.0 s) and rapid spin-down rate ($\dot{P} = 1$ ms/yr).

The magnetar model postulates that the dominant source of energy of magnetars is their intense magnetic field. The short bursts are produced as a result of outward propagation of Alfvén waves through the magnetosphere. The giant flare phenomenon is believed to be triggered by the sudden catastrophic reconfiguration of the magnetic field or by a fracture in the magnetically stressed crust (Thompson and Duncan 1995; Lyutikov 2003).

Rarely, SGRs produce enormously energetic giant flares with peak luminosities of the order of $10^{44} - 10^{46}$ erg s$^{-1}$, followed by pulsating tail lasting for hundreds of seconds. Only three hyper flares have been observed to date, with the first being the March 5, 1979 γ–ray hyper flare from SGR 0526-66 (Mazets et al. 1979). The two other giant flares were detected from SGR 1900+14 in August 1998 (Hurley et al. 1999; Feroci et al. 1999), and the most energetic giant flare ever recorded was detected from SGR 1806-20 in December 2004 (Hurley et al. 2005; Palmer et al. 2005). At the present time, there are 8 confirmed SGRs 0501+4516, 1900+14, 1806-20, 0525-66, 1627-41, 0418+5729, 1833-0832, and J1550-5418.

Early models for a neutron star with a fluid core and solid crust suggested that the modes most likely to be excited by a magnetar crustquake are torsional shear oscillations of the crust with a fundamental frequency ~ 30 Hz (Schumaker and Thorne 1983; McDermott et al 1998; Duncan 1998). The excited higher order harmonics will depend on the neutron star properties. After understanding the real nature of quasi-periodic X-ray oscillations, their observation will be very useful to put stringent constraints on neutron star masses and radii. In Chapter 4, we discuss the physical interpretation and theoretical modeling of QPOs in more detail.

### 3.2.1 Detected QPOs in the SGR 1806-20 giant flare

On 2004 December 27, SGR 1806-20 emitted its first recorded giant flare that was observed by a number of satellites as the most powerful giant flare from a magnetar. The giant flare reached peak luminosity of the order of $10^{46}$ erg s$^{-1}$ in the first 0.2 s causing a strong perturbation in the Earth's ionosphere and saturation of most satellite detectors (Hurley et al. 2005; Palmer et al. 2005). The initial spike was followed by ~ 400 s tail modulated at the 7.56 s spin period of the SGR. The detected QPOs are associated with hard emission component that dominates the overall energy emission about 170-200 s after the beginning of the giant flare. Using Rossi X-ray Timing Explorer (RXTE) Proportional Counter Array (PCA) observations in GoodXenon mode resulted in timing resolution of ~ 1 μs and spectral resolution covering the energy range 2 – 80 keV. Israel et al. (2005) discovered three X-ray QPOs at 18, 30, and 92 Hz in the pulsating decaying tail of the giant flare (see figure 3.1).

The 92 Hz QPO discovered in the decaying tail of the giant flare from SGR 1806-20 was strongly rotational phase dependent, appearing only for a fraction of the rotation cycle away from the main peak of the rotational pulse.

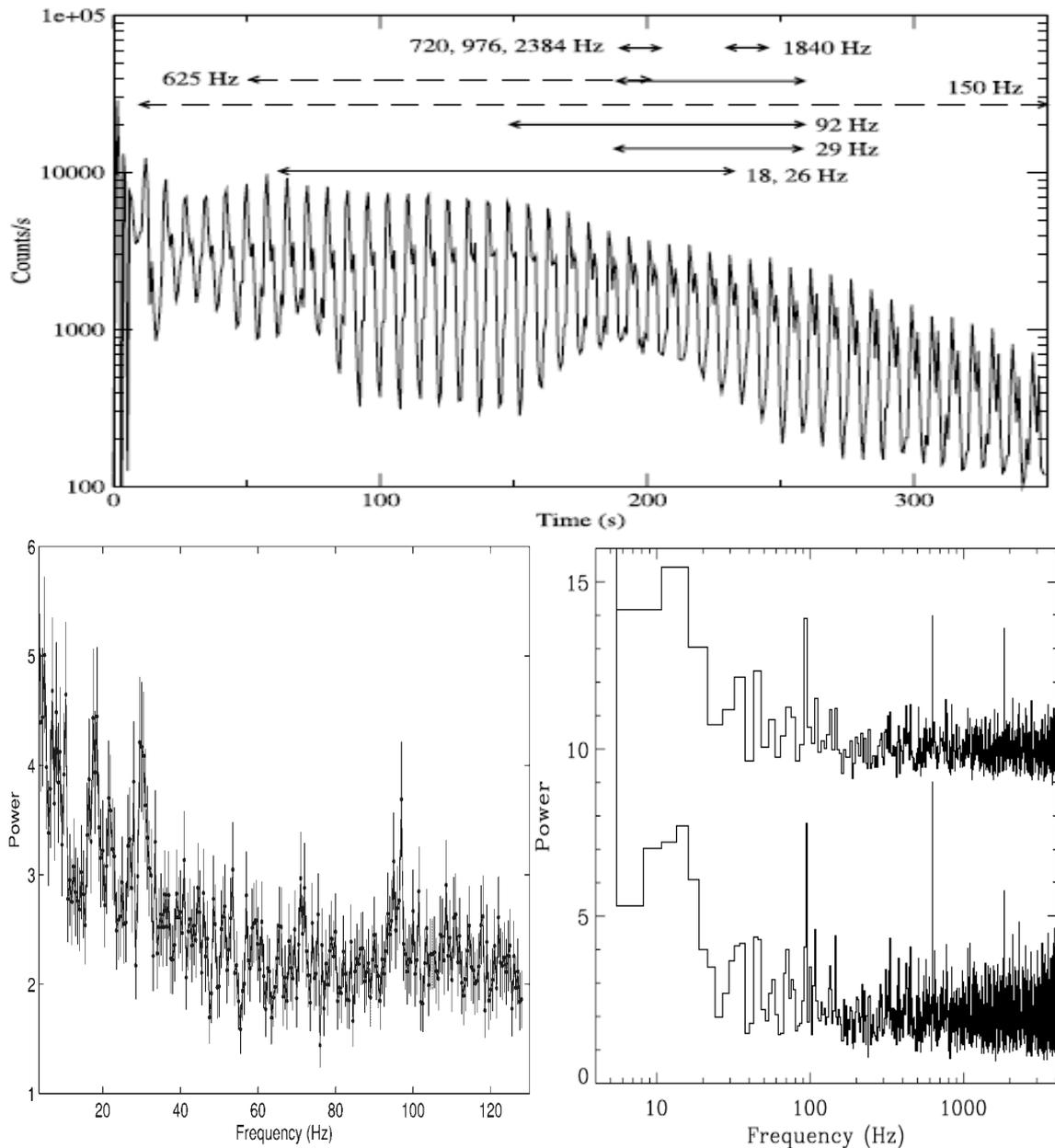

Figure 3.1 Light curve of the giant and the time periods where the different QPOs are detectable in either the RXTE or RHESSI data sets is shown (Upper panel). Power spectrum accumulated from data in the time interval 200–300 s (see lower left panel). Two low-frequency peaks at ~18 and ~30 Hz are visible, together with a significant peak at ~95 Hz. The lower right panel shows the average power spectrum. Two representations of the same power spectrum are shown, with only the frequency resolutions differing. The frequency bins are 2.667 Hz wide in the bottom trace and twice that in the top. Three frequencies are prominent in the top trace, 92, 625, and 1840 Hz. (Credit: Israel et al 2005 and Strohmayer & Watts 2006)

Strohmayer & Watts (2006) analyzed the SGR 1806-20 giant flare using Ramaty High Energy Solar Spectroscopic Imager (RHESSI), a solar-pointing satellite that covers a wider energy band than RXTE. RHESSI's detectors are split into front and rear segments. Their analysis confirmed the existence 18 and 92 Hz QPOs and reported two additional QPOs at 26 and 626 Hz. The detected QPO at 92 Hz was found at the same time and rotational phase found by Israel et al.

(2005). The additional sensitivity in RHESSI's dataset provided higher count rates and this revealed two broad peaks at 18 and 26 Hz at the same rotational phase as the 92 Hz QPO 50 – 200 s after the initial giant flare thus able confirming both the frequencies and the rotational phase dependence. Table 3.1 summarizes the detected QPOs in the SGR 1806-20 giant flare and their main properties.

Table 3.1 Summary of Properties for the detected QPOs in the tail of the SGR 1806-20 Giant Flare

| QPO Frequencies (Hz) | FWHM Width (Hz) | Duration (s) | Space Mission Satellite |
|---|---|---|---|
| 17.9 ± 0.1 | 1.90 ± 0.2 | 60 – 230 | *RHESSI* |
| 25.7 ± 0.1 | 3.0 ± 0.2 | 60 – 230 | *RHESSI* |
| 29.0 ± 0.4 | 4.1 ± 0.5 | 190 – 260 | *RXTE* |
| 92.5 ± 0.2 | $1.7^{+0.7}_{-0.4}$ | 150 – 260 | *RXTE* |
| 92.7 ± 0.1 | 2.3 ± 0.2 | 150 – 260 | *RHESSI* |
| 92.9 ± 0.2 | 2.4 ± 0.3 | 190 – 260 | *RXTE* |
| 150.3 ± 1.6 | 17 ± 5 | 10 – 350 | *RXTE* |
| 626.46 ± 0.02 | 0.8 ± 0.1 | 50 – 200 | *RHESSI* |
| 625.5 ± 0.2 | 1.8 ± 0.4 | 190 – 260 | *RXTE* |
| 1837 ± 0.8 | 4.7 ± 1.2 | 230 – 245 | *RXTE* |

Strohmayer & Watts (2006) re-analyzed the RXTE observations and confirmed the 26 and 626 Hz QPOs detected with RHESSI and reported additional QPOs at 150 and 1840 Hz. Two more QPOs at 720 and 2384 Hz were reported at a lower significance.

### 3.2.2 Detected QPOs in the SGR 1900+14 giant flare

In August 27, 1998, SGR 1900+14 emitted a giant flare with a peak luminosity ~ $10^{44}$ erg s$^{-1}$ that was detected by RXTE. Prompted by the intriguing discovery of QPOs in the giant flare from SGR 1806-20, Strohmayer & Watts (2005) studied each good interval separately, and discovered a strong transient 84 Hz QPO during a ~ 1 s interval about 60 s after the main flare. A pair of persistent QPOs at 53.5 and 155.5 Hz was strongly detected, while a third feature at 28 Hz is reported at a lower significance. All the reported QPOs were found at the same rotational phase as the 84 Hz signal and not detected at other rotational phases.

Strohmayer & Watts (2005) suggested that the phenomenology seen in SGR 1900+14 giant flare is similar to that reported by Israel et al. (2005) in the giant flare from SGR 1806-20, which implies a common origin. The discovered QPOs were identified as a sequence of toroidal modes associated with seismic vibrations in the neutron star crust (see Chapter 4 for details).

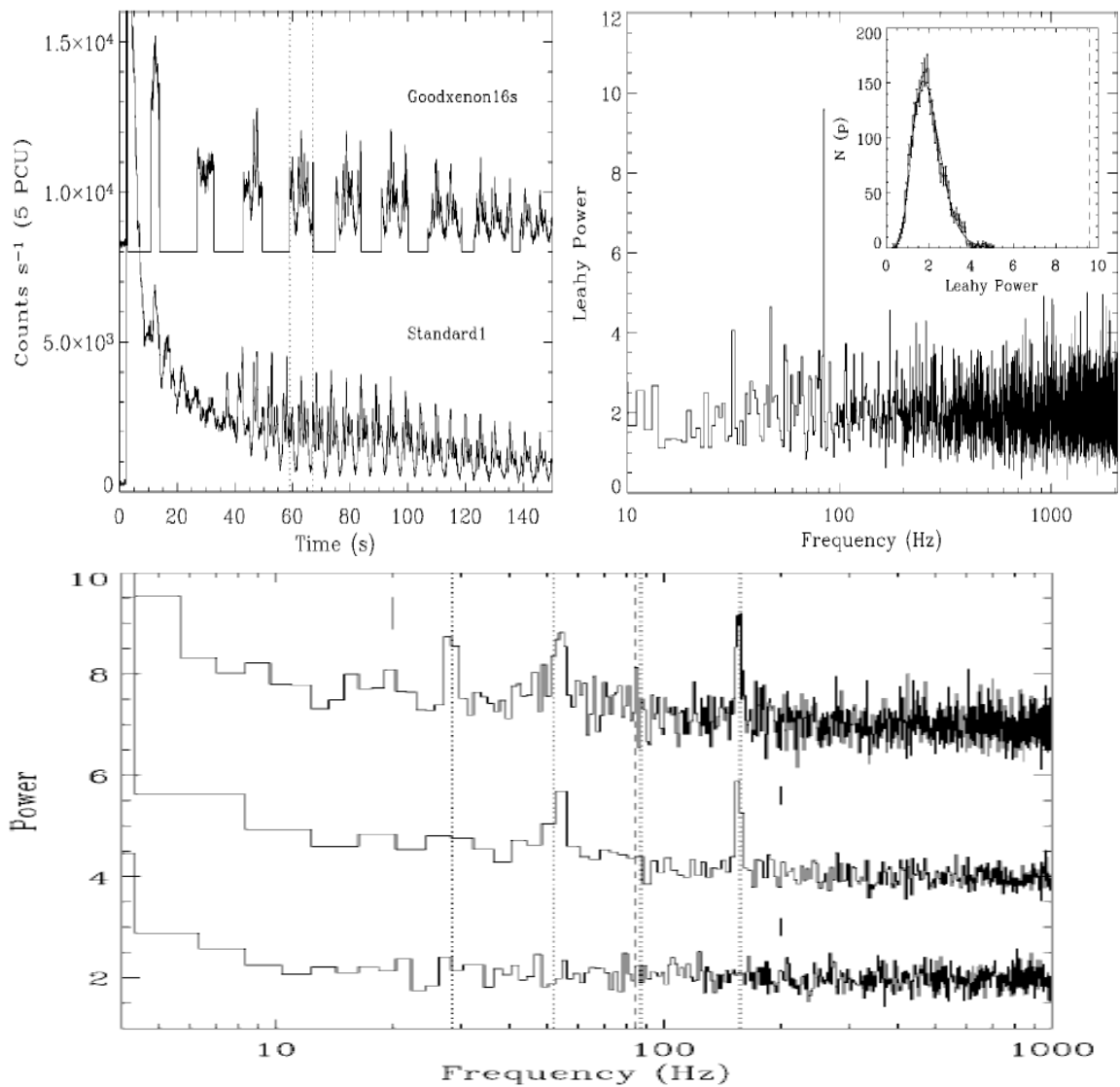

Figure 3.2 Light curves of the 1998 August 27 flare from 1900. The Goodxenon16s curve (top left panel) is plotted at 1/16 s resolution and is displaced vertically by 8000 s_1, for clarity. The upper right panel shows the power spectrum in the 10–2048 Hz band of the fourth good data interval from the Goodxenon16s data, showing the 84 Hz peak. The lower panel shows the average power spectra from 1.5 s intervals centered at different rotational phases (Credit: Strohmayer and Watts 2005).

## 3.3 Quasi-periodic oscillations in the recurrent emission

### 3.3.1 Introduction

"Neutron stars are great laboratories for the study of extreme physics". We'd love to be able to crack one open, but since that's probably not going to happen, observing the effects of a magnetar hyper flare on a neutron star is perhaps the next best thing." (Anna Watts)

However, due to the fact that High Energy Astrophysics is a "look but don't touch" science, the best thing we can do is to observe these extreme ultra-compact objects at a distance and from these observations we can make indirect inferences about the fundamental properties of these extreme objects. Using highly sophisticated space satellites we are able to have imaging, spectroscopy and timing information that can teach us about the nature and properties of neutron stars as well as other stellar objects. In our analysis we focused mostly on timing analysis to study the temporal properties of the source SGR 1806-20.

RXTE first observed SGR 1806-20 during November 1996. We obtained the PCA observations in 2-60 keV from RXTE public data archive (RXTE Data Archive). We reduced the data with HEAsoft 6.6 to obtain the burst light curves. The observations with the most bursts were taken in the event-mode configuration with time resolution $\Delta t \sim 125$ μs and 256 energy channels. All 5 proportional counter units (PCUs) of the PCA were fully functional and the photons from all PCUs and all detector anodes were combined to maximize the signal-to-noise ratio and the artificial event-mode time markers were removed. We screened out very faint bursts that do not have sufficient counts and very bright bursts that would cause significant pile-up and dead-time effects and selected 30 bursts for our timing analysis. We binned each burst at the smallest time resolution allowed by the data configuration (125 μs) and applied Fast Fourier Transformation (FFT) to the binned photon arrival times to obtain a power spectrum for each burst.

### 3.3.2 Timing and Fourier analysis

We study the temporal variations in the incoming flux from the neutron star and since the temperature range is considerably high, most of the emission falls in the X-ray and γ-ray region. The RXTE satellite PCA detectors detect the arrival time of each X-ray photon that hits the detector with high accuracy (usually microseconds). The accumulated count rate $c(t)$ is tabulated against the photon arrival time in order to produce the light curve of the emission (Count Rate vs. Time). The timing series $c(t)$ is transformed to the frequency domain using Fourier analysis to produce the normalized power density spectrum $P(\omega)$, where we have used Leahy normalization technique (Van der Klis 1989; Israel & Stella 1996):

$$P(\omega) = \frac{2}{N} \left| \int_0^T c(t) \exp(-i\omega t)\, dt \right|^2$$

Where $N$ is the total number of counts and $T$ is the duration of the observation. In that case, the pure Poisson noise will have a $\chi^2$ distribution and the probability that the power in a give time bin is greater than $P$ is $\exp(-P/2)$ and the fractional root mean squared variability at a certain frequency $\omega$ is $\left[ P(\omega)/N \right]^{1/2}$.

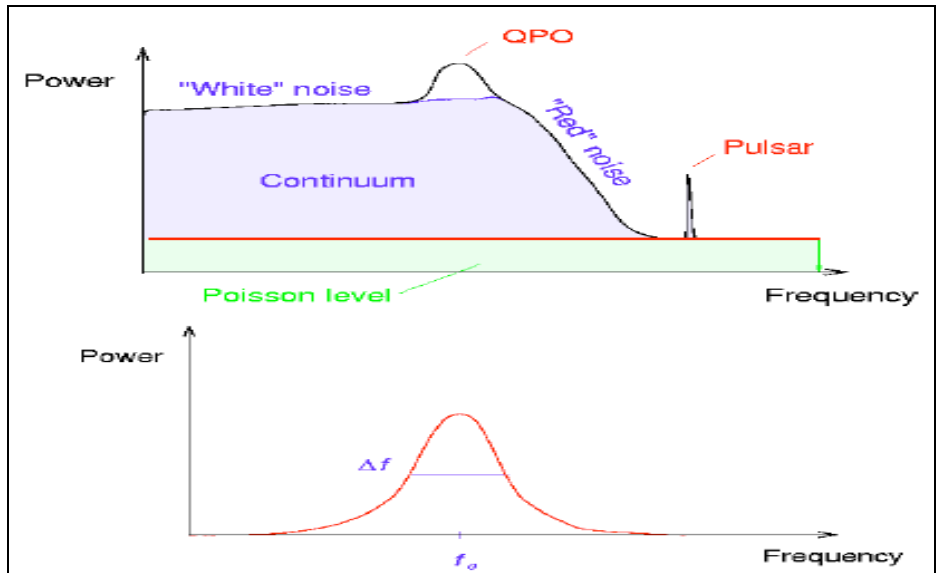

Figure 3.3 Illustrative schematic of the power spectrum showing the existence of a broad QPO (Credit: Craig Markwardt).

The timing analysis can give us more information on the dynamics of the objects, characteristic timescales, orbital periods, orbital evolution, broadband variability and quasi-periodic oscillations. A QPO is a messy oscillation (see figure 3.3) that can be due to intrinsic frequency variation, finite lifetime or amplitude modulation of the photon flux.

The frequency resolution $\Delta v = \dfrac{1}{\Delta t \times N_{bins}}$ ranged from 1.72 to 14.63 Hz for the bursts where we found evidence for QPOs, depending on the burst duration and the Nyquist frequency was $v_{Nyquist} = \dfrac{1}{2\Delta t} \sim 4096$ Hz. The power spectra were normalized using Leahy normalization method; we then searched each burst power spectrum for timing signals in the frequency ranges of the QPOs found in the 2004 giant flare. Candidate QPOs were fitted to a Lorentzian fitting function to calculate their properties, and their significance was investigated using Monte Carlo simulations.

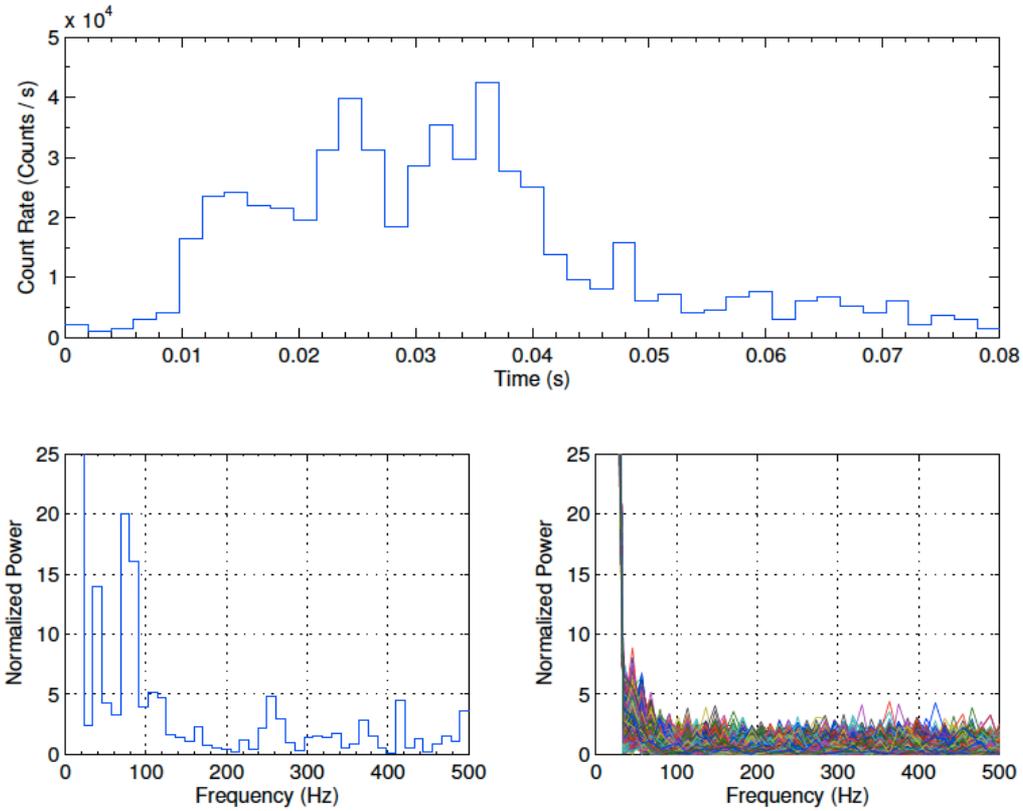

Figure 3.4 Light curve of the burst starting on 1996 Nov. 18 at 06:18:44.446 UTC (Upper panel). The lower left panel shows a QPO at 84 Hz in the burst power spectrum sampled at $\Delta v = 11.38$ Hz. The simulated power spectra are shown in the lower right panel.

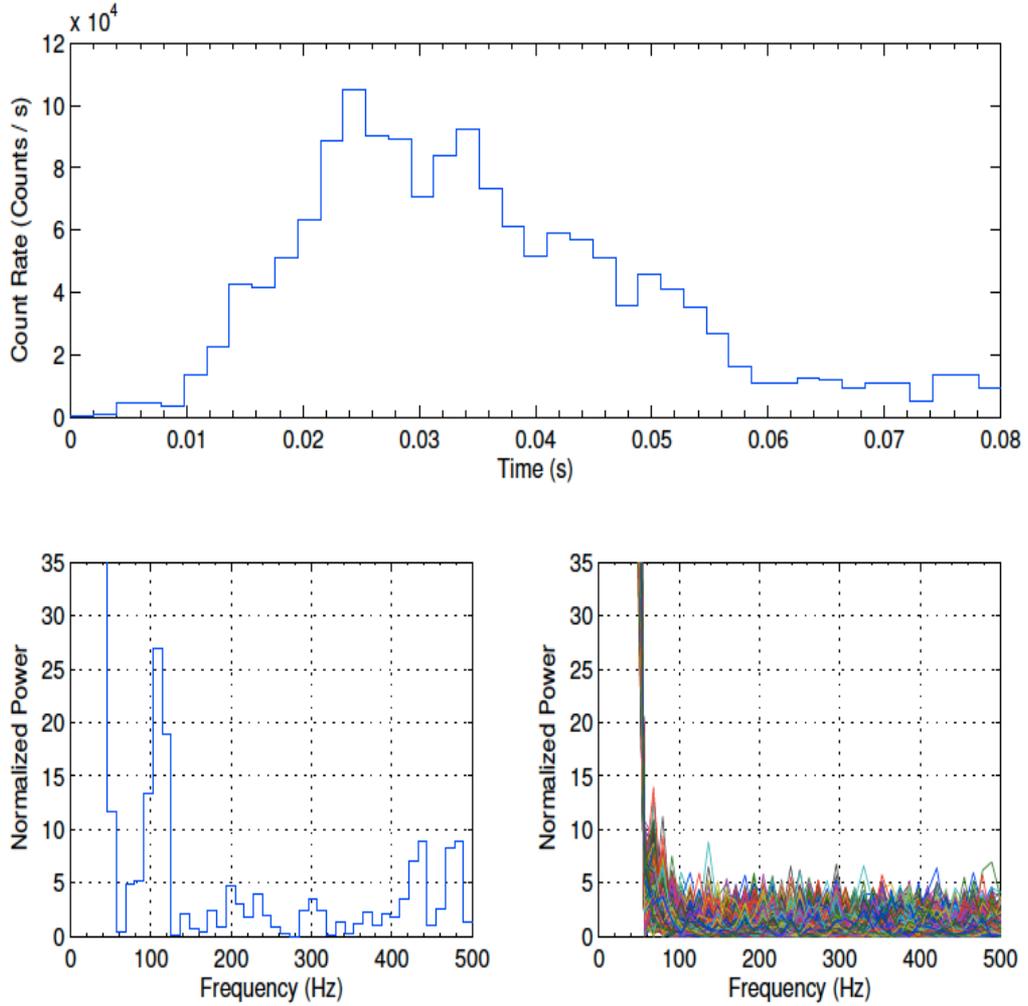

Figure 3.5 Light curve of the burst starting on 1996 Nov. 18 at 08:57:02.083 UTC (Upper panel). A QPO at ~ 103 Hz is evident in the burst power spectrum, sampled at Δν = 11.38 Hz (lower left panel). The simulated power spectra are shown in the lower right panel.

In a short, moderately bright burst (figure 3.4), we find a QPO candidate around 84 Hz with single trial null probability of $2.06 \times 10^{-4}$. The QPO has a centroid frequency $\nu_o = 84.31 \pm 0.66$ Hz Hz and a half width at half maximum (HWHM) $\sigma_\nu = 8.69 \pm 0.78$ Hz, corresponding to a coherence value $Q \equiv \nu_o / 2\sigma_\nu$ of 4.84. In another short burst with a gradual rise (figure 3.5), we detect a candidate QPO at 103 Hz at a null probability of $1.43 \times 10^{-6}$. This QPO has a centroid frequency $\nu_o = 103.44 \pm 0.42$ Hz, width $\sigma_\nu = 10.65 \pm 0.42$ Hz, and coherence value $Q \sim 5$.

In figure 3.6, we show a long (> 0.1 s), bright burst with a pair of QPO candidates at around 648 and 1096 Hz with a single trial probability of $2.33 \times 10^{-7}$ and $1.37 \times 10^{-5}$, respectively. Fitting each peak to a Lorentzian function yields $v_o = 648.5 \pm 0.15$ Hz, $\sigma_v = 2.5 \pm 0.09$ Hz, and $Q = 130$ for the first QPO and $v_o = 1095.88 \pm 0.21$ Hz, $\sigma_v = 2.88 \pm 0.17$ Hz, and $Q = 190$ for the second.

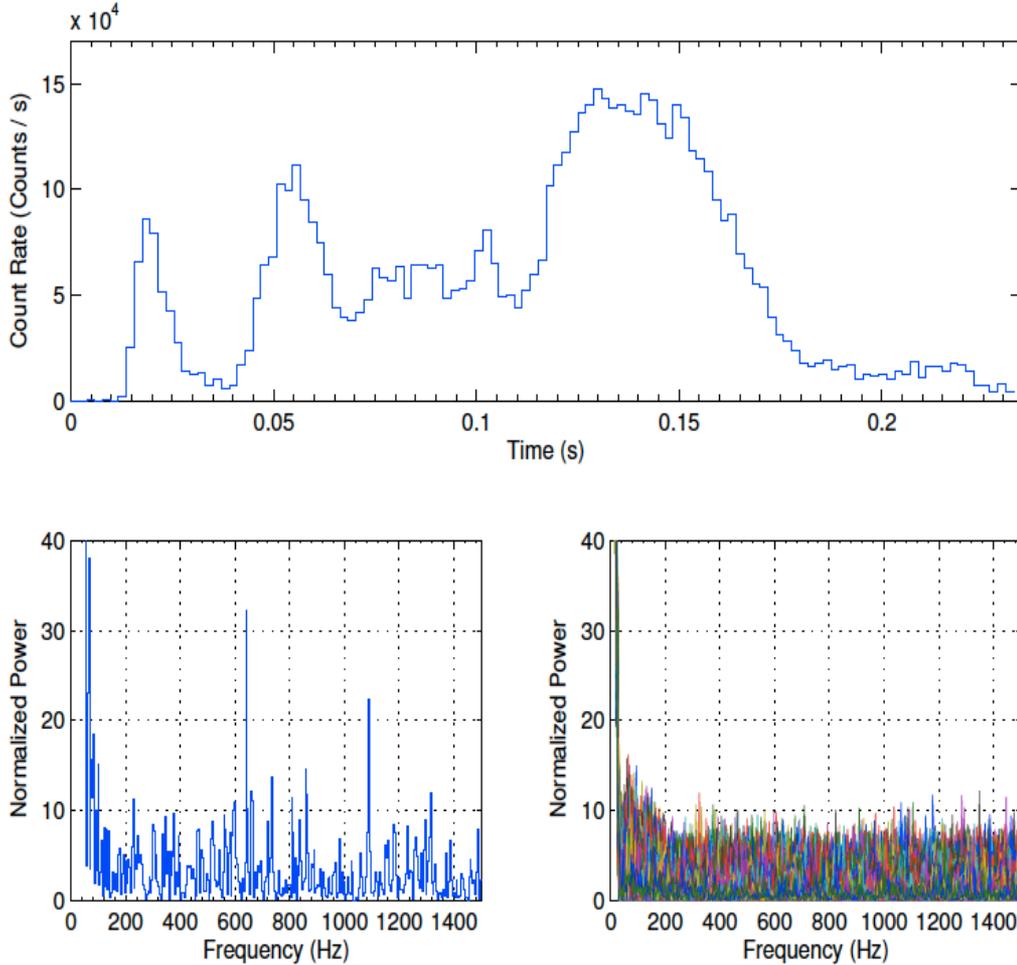

Figure 3.6 Light curve for the burst starting on 1996 Nov. 18 at 04:41:53.480 UTC (upper panel). The burst power spectrum ($\Delta v = 4.27$ Hz) shows two high-frequency QPOs at ~ 648 and 1095 Hz (lower left panel). The simulated power spectra are shown in the lower right panel.

In the other bursts we did not find significant QPOs at the previously detected frequencies. Extending our search to other frequencies, we detect two high frequency QPO candidates in the short burst shown in figure 3.7 at $1229.8 \pm 0.80$ and $3690.74 \pm 0.67$ Hz. Forming the null hypothesis results in a single trial probability of $3.4 \times 10^{-5}$ for each feature. We estimate that the 1230 and 3690 Hz features have $\sigma_v = 8.00 \pm 0.48$ and $8.37 \pm 0.79$ Hz respectively and corresponding coherence values $Q$ of 76 and 220. Another high frequency QPO candidate around 2785 Hz is found in the long, intense burst shown in figure 3.8, at a single trial probability of $7.96 \times 10^{-6}$. The QPO is centered at $v_o = 2785.36 \pm 0.20$ Hz and has a width $\sigma_v = 0.75 \pm 0.19$ Hz and coherence $Q = 1860$.

Since our strategy was to search for QPOs at the previously reported frequencies, the number of trials for the 84, 103, and 648 Hz QPOs takes into account the number of bursts and the number of frequency bins actually searched. This brings their null probabilities to $4.45 \times 10^{-3}$, $4.14 \times 10^{-5}$, and $3.85 \times 10^{-5}$, respectively. For the high frequency QPOs at 1096, 1230, 2785 and 3690 Hz, we had no prior knowledge of these frequencies and we must account for the total number of frequency bins in the power spectra as well as the number of bursts. This brings the chance probability to higher than $10^{-3}$, thus ruling out these QPOs on the basis of the null hypothesis alone.

Table 3.2  Summary of the properties for the most significant QPOs in the SGR 1806-20 recurrent emission during the November 18 1996 outburst.

| Event Date and Time (UTC) | QPO Frequencies (Hz) | HWHM Width (Hz) | Coherence Value $Q$ |
|---|---|---|---|
| Nov. 18, 1996 at 06:18:44.446 | 84.31 ± 0.66 | 8.69 ± 0.78 | 4.84848925 |
| Nov. 18, 1996 at 08:57:02.083 | 103.44 ± 0.43 | 10.65 ± 0.42 | 4.856160901 |
| Nov. 18, 1996 at 04:41:53.480 | 648.5 ± 0.15 | 2.50 ± 0.09 | 129.3609558 |
| Nov. 18, 1996 at 04:41:53.480 | 1095.89 ±0.21 | 2.88 ± 0.17 | 190.0031718 |
| Nov. 18, 1996 at 06:18:44.134 | 1229.80 ± 0.80 | 8.00 ± 0.48 | 76.82528025 |
| Nov. 18, 1996 at 12:18:01.487 | 2785.36 ± 0.21 | 0.75 ± 0.19 | 1860.195427 |
| Nov. 18, 1996 at 06:18:44.134 | 3690.74 ± 0.67 | 8.37 ± 0.80 | 220.5238019 |

### 3.3.3  Monte Carlo Simulation

A more cogent approach to assessing the statistical significance is through utilizing Monte Carlo simulations. Monte Carlo methods are a class of computational algorithms that rely on repeated random sampling to compute their results. Monte Carlo methods are often used in simulating physical and mathematical systems. Because of their reliance on repeated computation of random or pseudo-random numbers, these methods are most suited to calculation by a computer and tend to be used when it is unfeasible or impossible to compute an exact result with a deterministic algorithm. More broadly, it is the process of random generation of variables in a random process that is controlled and influenced by the boundary conditions. Monte Carlo simulation methods are especially useful in studying systems with a large number of coupled degrees of freedom

The Monte Carlo simulations were performed using MATLAB software, where for each of the above bursts, we generated a large number of simulated light curves at the same time resolution (125 μs) by fitting the burst light curve to a high-order polynomial function (see figure 3.9). We then determine the number of photons in each time bin by seeding a Poisson distributed random number generator whose mean value is the number of counts we obtain from the fitting function. The fitting function and the random generated Poisson distribution function are then superimposed and averaged to produce the simulated light curves. We used this approach to obtain 50,000 simulated light curves for each real burst and we generated a power spectrum for each simulated burst using the same technique we used to find the QPOs in the real bursts.

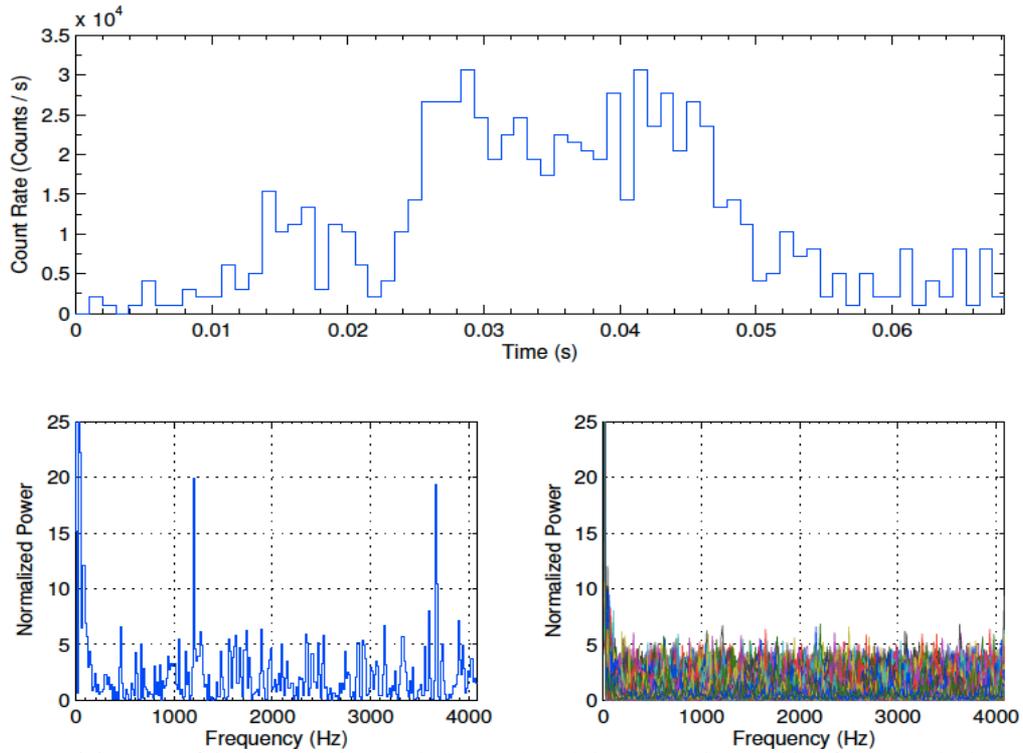

Figure 3.7 Light curve of the burst starting at 06:18:44.134 UTC (upper panel). Two high-frequency QPOs at ~ 1230 and 3690 Hz are evident in the power spectrum (Δν = 14.63 Hz) in the lower left panel. The simulated power spectra are shown in the lower right panel.

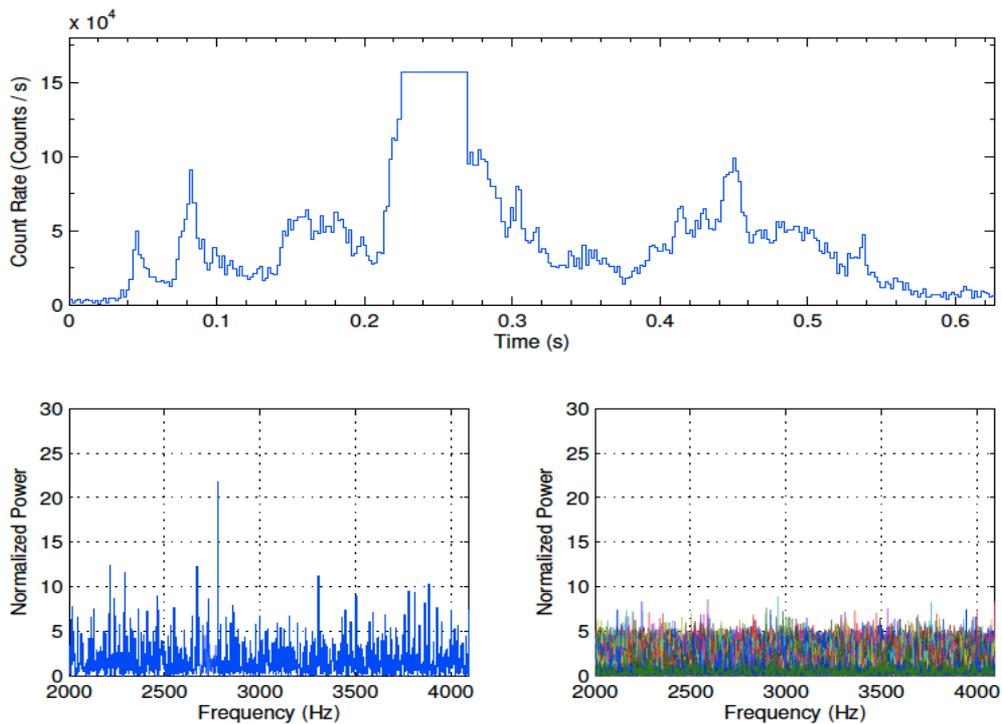

Figure 3.8 Light curve of the burst starting at 12:18:01.487 UTC (upper panel). The lower right panel shows a high-frequency power peak at ~ 2785 Hz in the burst power spectrum (Δν = 1.72 Hz). The simulated power spectra are shown in the lower right panel.

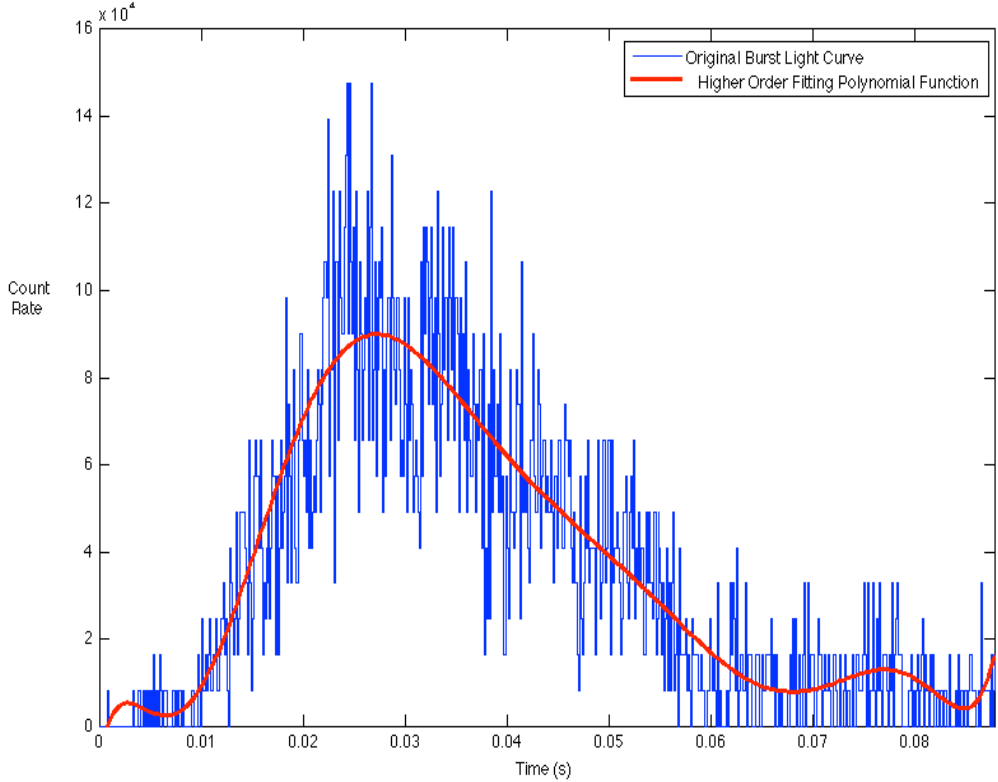

Figure 3.9 Plot of the original burst light curve at time resolution of ~ 125 µs and the high order polynomial fitting function.

In figures 3.4 – 3.8, the upper panel shows the light curve of each burst binned at 2 ms (with the exception of the burst in figure 3.7), while the lower panels show the power spectrum binned at the minimum resolution $\Delta t \sim 125$ µs and the lower right panels show the accumulated power spectra of the 50,000 simulated bursts[*2]. We find none to have shown a structure at or near the frequencies of the candidate QPOs with similar or higher power, yielding a null probability lower limit of $\leq 2 \times 10^{-5}$ corresponding to a confidence level interval $\geq 4.3\sigma$. The statistical significance arrived at from the two independent approaches above gives a secure robust detection for the 84, 103, and 648 Hz QPOs and a marginal significance for the kHz QPOs.

---

[*] The available hardware processing and memory resources to carry out the Monte-Carlo simulation restricted the number of simulated bursts.

# Chapter 4

# Theoretical Models and Interpretation of QPOs in Magnetars

## 4.1 Introduction

In this chapter, we compare our results against different theoretical models and discuss the possible physical interpretations of QPOs in magnetars. The detection of similar QPO Eigen frequencies in the giant flares from two different SGRs implies that we might be witnessing similar phenomenology and an analogous underlying physical mechanism in both sources.

## 4.2 Toroidal Seismic Modes Interpretation

Different theoretical models including torsional seismic modes, magnetic flux tubes oscillations, oscillatory modes within the magnetosphere and debris disk modes in accreting neutron stars were proposed to interpret the observed QPOs, their properties and implications on the neutron star physics. Existing models, which we discuss shortly, do have shortcomings. However, torsional seismic models seem promising for several reasons; (i) the detected QPO eigenfreqencies are consistent with the theoretical models and can be interpreted as a sequence of toroidal modes with varying spherical harmonic quantum number $l$ and $n=0$ for low frequency QPOs and higher order radial overtone modes $n > 0$ for high frequency QPOs. (ii) The magnetic field instability and reconfiguration powering the giant flares will most likely induce toroidal seismic vibrational modes causing fractures in the neutron star crust. (iii) The QPOs discovered in the giant flares event from the two sources revealed strong rotational phase dependence, which implies that the physical mechanism and fracture in the neutron star crust is associated with a particular area on the stellar surface. However, since we were investigating the typical recurrent bursts from SGR 1806-20 with typical duration less than 1 second, which made it difficult to investigate the phase dependence of each of the QPOs we report given that the period for SGR 1806 is ~ 8 seconds. (iv) The coupling of the magnetic field and the neutron star crust could cause the excited modes to modulate the X-ray flux. The present evidence implies that toroidal seismic models are successful in interpreting the physical origin of these QPOs, yet it is still not definitive since some of the low frequency QPOs detected (18 and 26 Hz) do not fit comfortably within this model unless other factors such as magnetic splitting and the coupling of the neutron star crust and core are taken into consideration.

Some papers that argue against global seismic vibrations as a viable physical mechanism for the generation of these QPOs mostly because of the crust-core coupling effect, where the vibrational motion of the crust would cause Alfvén waves to propagate into the core (see for example Levin 2006). Other models have been proposed as an alternative to seismic models including magnetic flux tube oscillations, modes of magnetosphere and modes of debris disk (Ma et al. 2009; Beloborodov & Thompson 2007; Wang et al. 2006). We confront and test the QPOs we detected in the recurrent burst emission and those previously detected in the giant flare against those models in this chapter.

### 4.2.1 Neutron Star seismic vibrational modes interpretation

In this section we discuss the implications of the QPOs in the light of toroidal mode models as a promising mechanism to interpret the observed frequencies and their properties.

Quasi-periodic oscillations were remarkably discovered from SGR 1806-20 during the giant flare on 2004 December 27 using RXTE/PCA observations at multiple frequencies including 92.5 Hz that was seen only over phases of the 7.56 second spin period yet no energy dependence was detected. Two other features at 18 and 30 Hz were also detected but with no specific phase dependence. Shortly after the detection of QPOs in the SGR 1806-20, detailed timing analysis of the August 27, 1998 giant flare from SGR 1900+14 using RXTE/PCA yielded the discovery and detection of QPOs at 84 Hz which appears to be associated with a particular rotational phase and two other QPOs were strongly detected at 53.5 Hz and 155.1 Hz and another broad feature at 28 Hz with lower significance. We summarize the QPO frequencies in the tails of the SGR 1806-20 and SGR 1900+14 giant flares in Table 4.1.

Table 4.1  Detected QPOs in the giant flare events from SGR 1806-20 and SGR 1900+14 and their toroidal mode assignment in terms of the quantum numbers *l* and *n*.

| SGR 1806-20 Detected QPO Frequencies (Hz) | SGR 1900+14 Detected QPO Frequencies (Hz) | Toroidal Mode Identification |
|---|---|---|
| 18 | | |
| 26 | | |
| 30 | 28 | $n=0, l=2$ |
| | 53 | $n=0, l=4$ |
| 92 | 84 | $n=0, l=6$ |
| 150 | | $n=0, l=10$ |
| | 155 | $n=0, l=11$ |
| 625 | | $n=1$ |
| 1840 | | $n=3$ |

Our analysis of the high-energy emission recurrent bursts from SGR 1806-20 shows interesting similarities to the results reported earlier in the giant flare phenomenon from SGR 1806-20 and SGR 1900+14 (Israel *et al.* 2005; Strohmayer and Watts 2005; Strohmayer and Watts 2006). The properties of the X-ray bursts that revealed the existence of QPO candidates are summarized in Table 4.2. We found three intriguing QPO candidates in SGR 1806-20 at 84, 103, and 648 Hz. The 84 Hz QPO candidate matches the feature found in SGR 1900+14 and lies within 9% of the 92.5 Hz QPO while the 103 Hz feature lies within 12% the 92.5 Hz feature in from the same source. The high frequency QPO candidate at 648 Hz is within 3.75% of the 625 Hz QPO discovered in 1806. The similarity in the detected QPOs is quite intriguing in its own right and may imply an analogous underlying physical mechanism. It was conjectured that toroidal oscillations of the neutron star might explain the observed frequencies (Israel *et al* 2005). We discuss our findings in context of this model in this section.

Table 4.2 Detected QPOs in the recurrent emission during the November 18, 1996 outburst from SGR 1806-20 and their toroidal mode assignment in terms of the quantum numbers *l* and *n*.

| Space Mission | SGR 1806-20 Detected QPO Frequencies (Hz) | Toroidal Mode Identification |
|---|---|---|
| RXTE | 84 | $n = 0$, $l = 6$ |
| | 103 | $n = 0$, $l = 8$ |
| | 648 | $n = 1$ |
| | 1096 | $n = 2$ |
| | 1230 | $n = 2$ |
| | 2785 | $n = 3$ |
| | 3690 | |

The magnetic instability powering the giant flares is believed to be associated with large-scale fracturing of the neutron star crust, which should excite shear toroidal vibrations of the neutron star (Thompson and Duncan 1995; Flowers and Ruderman 1977; Schwartz et al. 2005; Thompson and Duncan 2001). Earth seismologists regularly observe these global toroidal oscillation modes excited by crust fracturing in earthquakes as in the 2004 Sumatra-Andaman earthquake (Park et al 2005).

It has been suggested that the coupling between the magnetic field and the charged particles in the neutron star surface would trigger starquakes, causing global fracturing in the neutron star crust that result in the modulation of the X-ray light curve (Thompson and Duncan 1995, Thompson and Duncan 2001). These oscillations associated with toroidal vibrational modes of the neutron star are easier to excite than other possible oscillation mechanisms (Blaes et al 1989). The excited frequency modes depend on the neutron star mass and radius, crustal composition and magnetic field. The detection of these toroidal crustal modes has a great potential of revealing and probing the stellar structure, composition, magnetic field configuration and testing the neutron star equation of state (Duncan 1998; Hansen and Cioffi 1980; McDermott, van Horn, and Hansen 1988, Messios, Papadopoulous, and Stergioulas 2001; Piro 2005, Schumaker and Thorne 1983; Strohmayer 1991). Excitement of higher order $l$ modes effect was seen in the spectrum of Earth modes excited by the 2004 Sumatra-Andaman earthquake (Park et al 2005).

Duncan (1998) estimated the eigenfrequency of the fundamental toroidal mode $l=2$, $n=0$ for a non-rotating, non-magnetic neutron star, which we denote $v(_2t_0)$ to be

$$v(_2t_o) = 29.8 \frac{\sqrt{1.71 - 0.71 M_{1.4} R_{10}^{-2}}}{0.87 R_{10} + 0.13 M_{1.4} R_{10}^{-1}} \text{ Hz} \qquad (4.1)$$

Where $R_{10} \equiv R/10$ km and $M_{1.4} \equiv M/1.4 M_\odot$. If a twisted magnetic field is embedded throughout the neutron star crust, the eigenfrequencies need to be modified to

$$v = v_o \left[ 1 + \left( \frac{B}{B_\mu} \right)^2 \right]^{1/2} \text{ Hz} \qquad (4.2)$$

Where $v_0$ is the non-magnetic eigenfrequency, as in Eq. (4.1) for the fundamental mode $v(_2t_0)$. The higher order modes were shown to be scaled as $v(_lt_o) \propto \left[ l(l+1) \right]^{1/2}$ (Hansen and Cioffi 1980). The eigenfrequencies of the higher order $n=0$ toroidal modes were derived under the assumption that the magnetic tension boosts the field isotropically.

One must note that the degree to which the magnetic field modifies the eigenfrequency is highly dependent on the magnetic field configuration and other non-isotropic effects, which could alter this correction significantly (Piro 2005; Messios, Papadopoulous, and Stergioulas 2001). The derived eigenfrequencies of the higher order toroidal modes under the assumption of isotropic magnetic field are given by

$$\nu(_l t_o) = \nu(_2 t_o) \left[ \frac{l(l+1)}{6} \right]^{1/2} \left[ 1 + \left( \frac{B}{B_\mu} \right)^2 \right]^{1/2} \text{ Hz} \qquad (4.3)$$

Where the final factor is a magnetic correction factor, $B_\mu \approx 4 \times 10^{15} \rho_{14}^{0.4}$ Gauss and $\rho_{14} \sim 1$ is the density in the deep crust in units of $10^{14}$ g.cm$^{-3}$. The 92.5 Hz QPO and the weaker feature at 30.4 Hz discovered in SGR 1806-20 are suggested to be due to the $\nu(_7 t_0)$ and $\nu(_2 t_0)$ modes, respectively and the detection of the 626.5 Hz QPO was modeled as the $n = 1$ toroidal mode where the energy required to excite $n > 0$ modes is orders or magnitude larger than the energy required to excite an $n = 0$ mode which reflects the intensity of the flare.

Table 4.3 Neutron Star equations of state (Credit: Lattimer and Prakash 2001)

EQUATIONS OF STATE

| Symbol | Reference | Approach | Composition |
|---|---|---|---|
| FP | Friedman & Pandharipande (1981) | Variational | $np$ |
| PS | Pandharipande & Smith (1975) | Potential | $n\pi^0$ |
| WFF(1–3) | Wiringa, Fiks & Fabrocine (1988) | Variational | $np$ |
| AP(1–4) | Akmal & Pandharipande (1997) | Variational | $np$ |
| MS(1–3) | Müller & Serot (1996) | Field theoretical | $np$ |
| MPA(1–2) | Müther, Prakash, & Ainsworth (1987) | Dirac-Brueckner HF | $np$ |
| ENG | Engvik et al. (1996) | Dirac-Brueckner HF | $np$ |
| PAL(1–6) | Prakash et al. (1988) | Schematic potential | $np$ |
| GM(1–3) | Glendenning & Moszkowski (1991) | Field theoretical | $npH$ |
| GS(1–2) | Glendenning & Schaffner-Bielich (1999) | Field theoretical | $npK$ |
| PCL(1–2) | Prakash, Cooke, & Lattimer (1995) | Field theoretical | $npHQ$ |
| SQM(1–3) | Prakash et al. (1995) | Quark matter | $Q (u, d, s)$ |

NOTE.—"Approach" refers to the underlying theoretical technique. "Composition" refers to strongly interacting components ($n$ = neutron, $p$ = proton, H = hyperon, K = kaon, Q = quark); all models include leptonic contributions.

In SGR 1900+14 a sequence of modes of different $l$ is plausible in which the frequencies are due to the $l = 4$ (53 Hz), $l = 6$ (84 Hz), and $l = 11$ (155 Hz) modes with fundamental toroidal mode frequency $\nu(_2t_0) \approx 28$ Hz, lower than that for SGR 1806 yet consistent with the 28 Hz QPO detected in SGR 1900. The difference in the frequency of the fundamental $n=0$ modes for SGRs 1806-20 and 1900+14 can be attributed to a difference in the properties such as mass, radius or magnetic field strength of the two magnetars for a given equation of state (EOS).

The composition of a neutron star is mostly dominated by the nature of strong interactions, which are not well understood in dense matter. Most models that have been investigated can be conveniently grouped into three broad categories: non-relativistic potential models, relativistic field theoretical models, and relativistic Dirac-Brueckner-Hartree-Fock models. In each of these approaches, the presence of additional softening components such as hyperons, Bose condensates, or quark matter can be incorporated. In high-energy astrophysics, the Tolman-Oppenheimer-Volkoff (TOV) equation constrains the structure of a spherically symmetric body of isotropic material which is in static gravitational equilibrium, as modeled by general relativity. The equation reads

$$\frac{dP(r)}{dr} = -\frac{G}{r^2}\left[\rho(r) + \frac{P(r)}{c^2}\right]\left[M(r) + 4\pi r^3 \frac{P(r)}{c^2}\right]\left[1 - \frac{2GM(r)}{c^2 r}\right]^{-1} \tag{4.4}$$

where $\rho(r_0)$ and $P(r_0)$ are the density and pressure, respectively at $r = r_0$.

The above equation is derived by solving the Einstein equations for a general time-invariant, spherically symmetric metric given by

$$ds^2 = e^{\nu(r)}c^2 dt^2 - \left(1 - 2GM(r)/rc^2\right)^{-1} dr^2 - r^2\left(d\theta^2 + \sin^2\theta d\phi^2\right)$$

where $\nu(r)$ is determined by the constraint $\qquad(4.5)$

$$\frac{d\nu(r)}{dr} = -\left(\frac{2}{P(r) + \rho(r)c^2}\right)\frac{dP(r)}{dr}$$

From the TOV equation we obtain the neutron star matter pressure in terms of the mass and radius. The TOV limit sets an upper bound to the mass of neutron stars composed on neutron degenerate matter. Further theoretical limits from General relativity and causality on the maximum mass and minimum period of the neutron star allows us to set constraints on the equation of state. The different equations of state for the neutron star are summarized in Table 4.3, where the theoretical technique is varied as well as the composition structure of strongly interacting components in the neutron star. Figure (4.1) shows four different EOS discussed in Lattimer and Prakash (2001), where if we assume similar magnetic field strength, the masses of the two magnetars must differ by more than 0.2 $M_\odot$. The masses of radio pulsars have been found to have a narrow Gaussian distribution profile, $M = (1.35 \pm 0.04)\ M_\odot$. It is more likely that the two SGRs have similar masses but different magnetic field strengths. If we consider both stars to have a mass of the order 1.35 $M_\odot$, therefore in this case neither the softest nor the stiffest EOS will account for the magnetic field strength. The largest masses and stiffest EOS (MS0) require magnetic field strengths that are far higher than those inferred from timing analysis (Woods et al. 2002), while the softest EOS (WFF1) require extremely high masses that will not account for the calculated fundamental frequencies for both SGRs 1806-20 and 1900+14. The intermediately stiff EOSs AP3 and AP4 can account for the observed frequencies and yield magnetic field strengths that are fairly consistent with those obtained from timing studies (Woods and Thompson 2005).

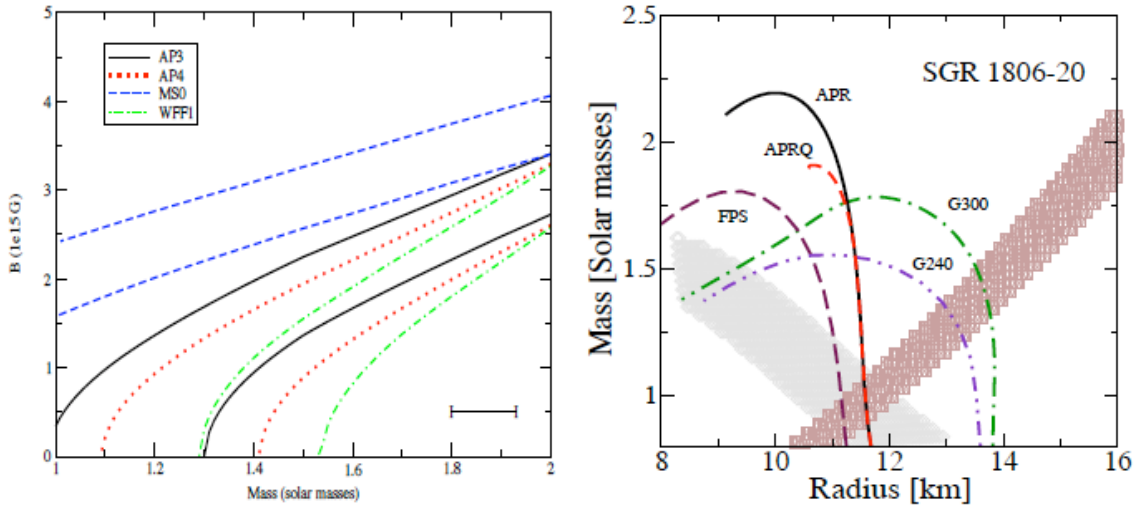

Figure 4.1. The left panel shows the constraints on the neutron star structure for 4 different equations of state, where for each EOS the upper line is for SGR 1806-20 and the lower line is for SGR 1900+14 (Credit: Strohmayer and Watts 2005). The allowed regions based on $n = 0$ and $n = 1$ mode identifications for SGR 1806-20 are shown in the right panel, where the shaded region with decreasing mass versus radius denotes the $n = 0$ constraints (Credit: Samuelsson and Andersson 2007).

We discuss our findings and results for the detection of QPOs in the recurrent burst emission from SGR 1806-20 in context of the toroidal mode frequencies. Using equations (4.1) and (4.3), we assign the observed low frequencies a sequence of toroidal modes of increasing $l$ and $n = 0$, where the QPOs at 84 and 103 Hz correspond to $\nu(_6t_0)$ and $\nu(_8t_0)$ respectively. Ever since the detection of the magnetar QPOs there has been an incredible boost in the theoretical efforts in this area to interpret the observed frequencies and their properties. We can see clearly from the above discussion that the excited torsional shear modes depend on the global structure of the neutron star and hence the equation of state (Lattimer and Prakash 2001 and 2004). The state of theoretical modeling have been refined and improved to include other factors and effects have in numerous papers addressing issues that include the crust/core coupling (Glampedakis et al 2006); crustal composition (Piro 2005); gravitational red shift; boundary layer physics (Kinney and Mendell 2003); elastic properties at the base of the crust (Pethick and Potekin 1998); magnetic field geometry and the effect of the strong magnetic field pressure on the tension in the star crust which would result in the alteration of the mode frequencies non-isotropically (Duncan 1998; Messios, Papadopolous, and Stergioulas 2001).

### 4.2.2 Recent developments in Toroidal seismic vibrational models

Shaisultanov and Eichler (2009) recently studied the effect of magnetic field on the frequencies of the toroidal oscillations in the QPOs detected in the SGR 1806-20 giant flare. It was suggested that we might be witnessing magnetic splitting of degenerate toroidal modes caused by the extremely high crustal magnetic field of the order $10^{15}$ Gauss. The SGR 1806-20 giant flare power spectrum contained a number of peaks spanning the 78 - 103 Hz frequency range that could be attributed to splitting. The two QPOs we report at 84 and 103 Hz may be degenerate modes induced as a result of the magnetic splitting. The change or shift in the frequencies of the observed QPOs between the 2004 giant flare and the 1996 observations could be a result of the decay in magnetic field. If we assume that the 103 and 92 Hz QPOs are attributed to $l = 7$ mode, we obtain a magnetic field strength of $2 \times 10^{15}$ and $3.5 \times 10^{14}$ G respectively, which reflects the decay in the magnetic field over a period of 8 years between the two observations. Within the perturbation theory framework and assuming a uniform magnetic field $B_o$ in the $z$ – direction and uniform shear modulus $\mu$, the eigenfrequencies were derived to lowest order under the condition that $\frac{B_o^2}{8\pi\mu} \ll 1$.

In a recent paper, a general relativistic formulation using Cowling approximation (no correction for magnetic field effects) was devised to explore torsional modes taking into consideration the gravitational red shift and the crust thickness yielded a set of revised estimates (Samuelsson and Andersson 2007). From the numerical calculations, it was estimated that the frequency modes for the $n = 0$ and $n > 0$ modes to be

$$\nu(_l t_0) \approx 27.65 \frac{1}{R_{10}} \sqrt{\frac{(l-1)(l+2)}{2}} \frac{\sqrt{\beta_*(1.705-0.705\beta_*)(0.1055+0.8945\beta_*)}}{1.0331\beta_* - 0.0331} \text{ Hz}, \quad (n=0) \qquad (4.6)$$

$$\nu(_l t_n) \approx 1107.3 \frac{n(0.1055+0.84945\beta_*)}{R_{10}} \frac{\beta_*}{1.0166\beta_* - 0.0166} \text{ Hz}, \quad (n>0) \qquad (4.7)$$

The above expressions are scaled differently than the $[l(l+1)]^{1/2}$ scaling in previous estimates by Duncan (1998) and Piro (2005). Using the general relativistic formulation with Cowling approximation yields equation (4.6) which may lead to different mode assignments than those obtained using equation (4.3) as in the case of the 92.5 Hz QPO that is assigned an $l = 6$ and $n = 0$ mode in this model instead of $l = 7$ and $n = 0$ based on previous estimates.

Interestingly the 84 Hz QPO candidate matches the feature found in SGR 1900+14 and lies within 9% of the 92.5 Hz QPO while the 103 Hz feature lies within 12% the 92.5 Hz feature in from the same source. The 84 Hz QPO we report in the recurrent emission and the 92 Hz QPO that was detected in the giant flare are attributed to the same toroidal ($l = 6$, $n = 0$) mode. The high frequency QPO at 648 Hz that lies within 3.5% from the 625 Hz QPO previously discovered in the giant flare event from the same source. The high frequency feature we report in SGR 1806-20 at 648 Hz may be interpreted in terms of $n = 1$ toroidal crust mode for which $l$ is not greatly constrained. During the giant flare from SGR 1806, RXTE mission also detected a QPO at 1840 Hz with fractional amplitude too low to be detected in RHESSI. The torsional shear mode identified for the 1840 Hz is $n=3$ where again $l$ is not constrained. For the extremely high eigenfrequencies at 1096, 1230, 2785 and 3690 Hz we use the general relativistic formulation to assign a set of higher radial overtone modes ($n > 0$) summarized in Table (4.2). The $n > 1$ torsional shear modes identification and seismic vibrations in the neutron star crust might be a more plausible interpretation for the high frequency QPOs. Piro (2005) calculated the torsional shear mode properties assuming a constant vertical magnetic field $\left(B = B_o \hat{z}\right)$ and including the gravitational red shift effect from the neutron star surface. In this model the estimated observed frequencies for the $n = 0$ shear modes are given by

$$\left.\frac{\omega_{obs}}{2\pi}\right|_{n=0} = 28.8 \rho_{14}^{0.6} \left(\frac{Z}{38}\right)\left(\frac{302}{A}\right)^{2/3}\left(\frac{1-X_n}{0.25}\right)^{2/3}\left[\frac{l(l+1)}{6}\right]^{1/2} R_{12}^{-1}\left(1.53 - 0.53\frac{M_{1.4}}{R_{12}}\right)^{1/2} \text{ Hz} \qquad (4.8)$$

Where $R_{12} \equiv R/12$ km, $X_n$ is the fraction of neutrons, $Z$ is the charge per ion and $A$ is the number of nucleons per ion. From equation (4.8), the observed frequencies are independent of the magnetic field strength, which is mainly because of the simple magnetic field geometry assumption. This model sets an upper limit on the magnetic field strength (critical magnetic field $B_{crit}$) for which the above equation holds. When $B \geq B_{crit}$, the frequency estimates are no longer valid. Given a shear wave speed $\upsilon_s = \sqrt{\frac{\mu}{\rho}}$ and Alfvén wave speed $\upsilon_A = \frac{B}{\sqrt{4\pi\rho}}$, we estimate the magnetic field effect at an angle $\alpha$ from the vertical direction such that $\upsilon_s = \upsilon_A \sin\alpha$ and from the definition of the Alfvén wave speed we get the critical magnetic field strength to be

$$B_{crit} = 3.8\times10^{15}\frac{\rho_{14}^{2/3}}{\sin\alpha}\left(\frac{Z}{38}\right)\left(\frac{302}{A}\right)^{2/3}\left(\frac{1-X_n}{0.25}\right)^{2/3} \text{ Gauss} \qquad (4.9)$$

Figure 4.2 asserts that the $n=0$ modes are independent of the magnetic field and scale as $[l(l+1)]^{1/2}$ based on Piro's analytical and numerical results whereas the $n > 0$ modes exhibits degeneracy in the quantum number $l$ and for low magnetic field strength we have $\omega \propto B$. The theoretical model predicts that the frequency range of $\sim 600 - 3000$ Hz seem to be quite an interesting region for conducting searches for the $n > 0$ modes that would also be very useful for constraining the neutron star equation of state. It is quite interesting that some of the QPOs we detected in the SGR 1806-20 recurrent burst emission (648, 1096, 1230 and 2785 Hz) fit within the theoretical predictions and can further enhance our chances of getting better constraints on the crustal properties of the neutron star.

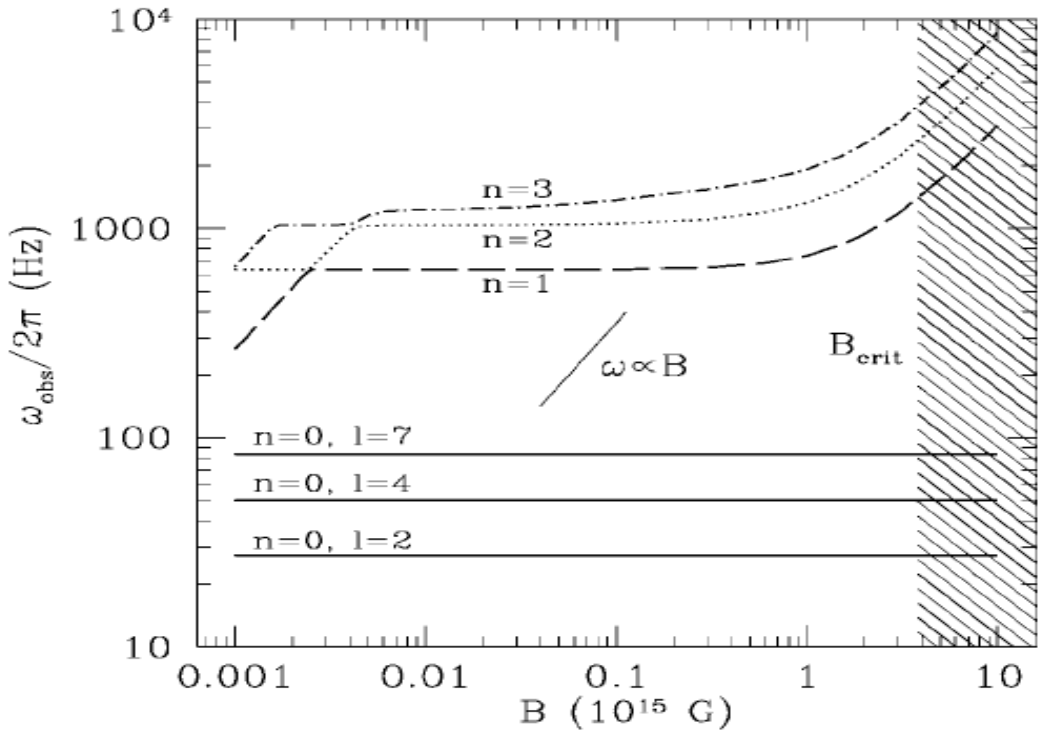

Figure 4.2 Giant flare observed QPO frequencies of the $n=0, 1, 2$ and 3 toroidal shear modes as a function of the magnetic field $B$. For $B \geq B_{crit}$ (shaded region) the frequency estimates cannot be trusted particularly if the magnetic field has a non vertical component (Credit: Piro 2005)

Watts and Strohmayer (2007) argued that if the $n=1$ toroidal crust modes interpretation is correct, this will add more constraints on the nuclear equation of state and also provide further insight on the neutron star crust thickness. In the limit of a thin crust ($\Delta R \gg R$), it can be shown using the perturbative calculations done by Hansen and Cioffi (1980) that the ratio of the neutron star crust thickness $\Delta R$ to the stellar radius $R$ is directly proportional to the fundamental mode frequency $\nu(_2t_0)$ to the first radial overtone ($n=1$) toroidal mode $\nu(_lt_1)$ as follows.

$$\frac{\nu(_lt_0)}{\nu(_lt_n)} = \frac{[l(l+1)]^{1/2}}{3n}\frac{\Delta R}{R} \tag{4.10}$$

For the SGR 1806-20, we have $\nu(_2t_0)$ = 30 Hz and $\nu(_lt_1)$ = 625 Hz for the giant flare and $\nu(_lt_1)$ = 648 Hz. Substituting these values into equation (4.8) yields $\Delta R/R \sim 0.06$. More recent estimates that do not assume a constant shear speed give a crust thickness of 0.1 – 0.12 times the stellar radius. More detailed modeling is refining this estimate after including gravitational red shift and crust thickness and further efforts are required to include the crust/core coupling will lead to revised estimates. Measuring and estimating the neutron star crust thickness gives additional constraints on the EOS and stellar parameters since these quantities depend on the global structure of the neutron star and hence the equation of state of matter in the deep core. The direct estimate and measurement of the neutron star crust is of great importance for the self-bound strange star models, which are hypothetical compact objects conjectured by Witten (1984). Strange stars are assumed to be composed entirely of stable strange quark matter with predicted properties that are very similar to those of neutron stars except for a major difference where the strange stars are predicted to have a much thinner crust thickness (Alcock et al. 1986; Jaikumar et al. 2006). Recent investigation of the torsional shear mode frequencies for the strange star crust models were found to be substantially different from those of found from observational results in neutron stars, thus ruling out the self-bound strange stars (Watts and Reddy 2007).

Several effects need to be taken into consideration including crust/core coupling, magnetic field geometry, viscous and elastic properties at the base of the crust and non-homogeneity in the crust among others for more accurate and refined models. It was predicted by (Prendergast, 1956) (1956) that a stable field inside a star may consist of a polar dipole component stabilized by a toroidal component of comparable strength, therefore the assumption of an isotropic magnetic field is inadequate since magnetars have strong dipole field components and internal poloidal-toroidal torus fields as demonstrated using numerical Magnetohydrodynamics (MHD) calculations by Braithwasite and Spruit (2006).

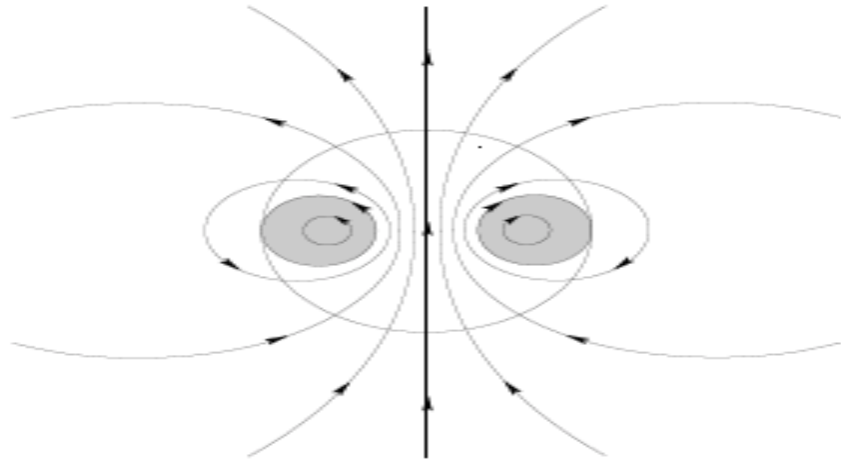

Figure 4.3. The stable linked poloidal-toroidal torus field. Poloidal field lines are drawn; the shaded areas represent the azimuthal field component (Credit: Braithwasite and Spruit 2006)

Over the years a number of papers have addressed issues including the magnetic field geometry, crust-fluid core coupling due to the presence of the magnetic field that modifies the boundary conditions at the interface (Carroll et al. 1986; Messios, Papadopoulous, and Stergioulas 2001 and Glampedakis et al. 2006). An attempt to approach this problem was made by (Glampedakis et al. 2006), where they used a toy-model to compute the global MHD modes of a neutron star taking into account the magnetic coupling between the elastic crust and the fluid core for a uniform field using simple parallel-plane slab geometry where the fluid core is sandwiched between two elastic slabs of crust. The results from this simple model were highly suggestive as they shed light on the effect of inclusion of coupling on calculating the global magneto-elastic modes that provided us with a spectrum of global modes both in the crust and the core with frequencies that are very similar to the eigenfrequencies of QPOs detected in the SGR 1806-20 and provides a natural explanation for the extremely low frequency QPOs at 18 and 26 Hz. This model also supported seismic vibrational mode interpretation and suggests the idea that the system will naturally excite modes, which minimize the transfer of energy into the core and the modes that will survive are those for which the coupling to the neutron star core is feeble. Until present time, the global seismic vibration model remains to be the most promising mechanism for the interpretation of QPOs in magnetars yet we discuss other suggested mechanisms and models that argue against global seismic modes interpretation below.

### 4.2.3 Arguments against the seismic vibrational modes interpretation

Levin (2006) argued against the global seismic vibrations as a viable interpretation for the QPOs in magnetars and suggested that the QPOs could be associated with MHD elastic modes of the neutron star. The model considered a perfectly conducting incompressible fluid sandwiched in a box, with top and bottom plates also being perfect conductors (see Figure 4.4). The magnetic field $\vec{B}(y)$ is vertically directed and is a function of $y$ only, threading both top and bottom plates. Solving the Lagrangian displacement equation of motion in the $z$-direction, Levin obtained a continuous spectrum of eigenfrequency modes rather than a discrete spectrum:

$$\frac{\partial^2 \zeta}{\partial t^2} = c^2(y)\frac{\partial^2 \zeta}{\partial x^2} - \gamma \frac{\partial \zeta}{\partial t} \tag{4.11}$$

where $c(y) = \sqrt{T(y)/\rho(y)}$ and $\gamma$ is a small damping constant; here $T = B^2/4\pi$ is the magnetic tension (In a superconducting fluid $T = BB_{crit}/4\pi$). The eigenfunctions in the $x$ and $y$ are given by:

$$\zeta_{nyo}(x,y) = \sin(n\pi x/l_x)\delta(y - y_o)\exp(i\omega_{nyo}t) \tag{4.12}$$

with eigenfrequencies

$$\omega_{nyo} = \pm\sqrt{(n\pi/l_x)^2 T(y_o)/\rho(y_o) - \gamma^2/4} + i\gamma/2 \tag{4.13}$$

Where $l_x$ is the box height, $n$ is an integer and $y_o$ is the $y$-coordinate where the eigenfunction is localized. This yields a continuous spectrum of eigenfrequencies.

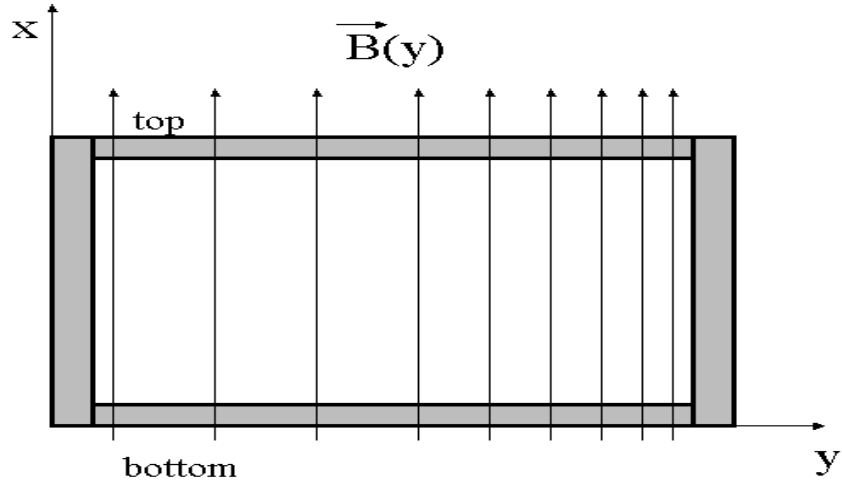

Figure 4.4. Toy model of the magnetar structure proposed by Levin (Credit: Levin 2006)

The toy model employed by Levin assumed a rather simplified model for the geometry of the neutron star and the boundary conditions used were inadequate and the resulting continuous spectrum is an artifact due to the assumption of an infinitely long slab extended in the *z*-direction. Watts and Strohmayer (2007) offered a detailed critique and falsified Levin's conclusions except for the crust/core coupling effect which needs to be well accounted for in future models.

## 4.3   Alternative Interpretations of QPOs in Magnetars

### 4.3.1   Magnetic flux tubes and Magnetospheric oscillation models interpretation

Another recent interpretation of the low frequency QPOs that could be modeled and well accounted for is using the standing sausage mode oscillations of flux tubes in the magnetar corona (Ma, Li, and Chen 2008). It is assumed that part of the plasma ejected during the giant flares is trapped by the magnetic fields and then forms magnetic flux tube structures similar to what is seen in the solar corona. However, the proposed model accommodates only for low frequency QPOs, while the very high frequency candidates we report at 648, 1096, 1230, 2785 and 3690 Hz in SGR 1806-20 are difficult to explain in the tube oscillation model. For the very high frequency QPOs, Beloborodov and Thompson (2007) have suggested that the inner electric current in magnetar corona will induce very high frequency QPOs ~ 10 kHz where the corona self organizes itself into quasi steady state. Such high frequency QPOs remain to be detected while the lower frequency oscillations would involve the outer corona where the emission in this region is not intense enough to excite the observed QPOs.

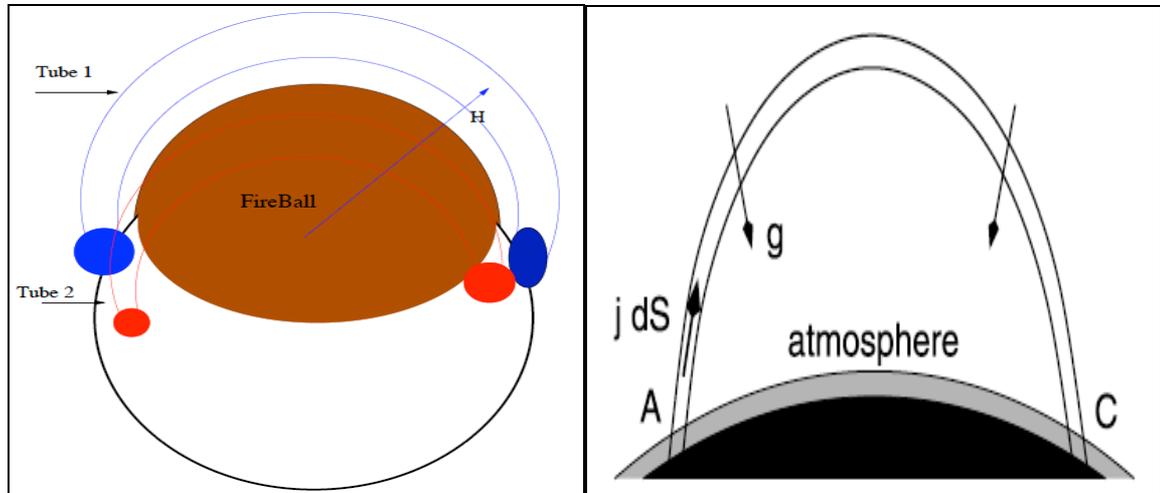

Figure 4.5 Illustrative schematic of the flux tubes in the SGR giant flare is shown in the left panel (After Ma et al. 2008). The right panel shows a schematic picture of a current-carrying, closed magnetic flux tube, anchored to the star's surface. The current is initiated by a starquake that twists one (or both) foot-points of the tube (After Beloborodov and Thompson 2007).

The QPOs we report are detected in short ($< 0.1$ s), relatively dim bursts and in long ($> 0.1$ s), bright bursts that would produce different signatures in the magnetosphere. Since we detect QPOs in weak bursts that may not produce a fireball or be of magnetospheric origin, this casts doubts on magnetospheric oscillation models as an origin for the QPOs in magnetars.

### 4.3.2   Modes of an accreting debris disk

The discovery of a debris disk around the Anomalous X-ray Pulsar 4U 0142+61 by Wang et al. (2006) led to the suggestion that the magnetar QPOs could be accounted for by similar phenomena since many accreting neutron stars show kHz QPOs that are believed to originate in the accretion disk (Van der Klis 2006). The reported QPOs are detected during the bright X-ray emission of the bursts and are not seen in the decaying phase, arguing against arising from a passive debris disk around the neutron star. Furthermore, this interpretation is mainly ruled out by the fact that we have no evidence for the existence of a fallback passive debris disk around SGRs. Watts and Strohmayer (2007) pointed out that for neutron stars that exhibit extremely high frequency QPOs, the inner disk radius of the debris disk is comparable to that of the neutron star. In the case of magnetars, the inner disk radius of the debris disk would have to be several solar radii to induce the observed magnetar QPOs, which rules out the modes of a debris disk mechanism as a possible interpretation for the magnetar QPOs.

## 4.4 Theoretical Challenges

The frequencies of the QPOs are in good agreement, for the most part, with models of torsional shear modes of neutron star crusts – in accordance with the early theoretical suggestions. The models used to make these identifications were based on early calculations that assumed free slip of the solid crust over the fluid neutron star core (Hansen and Cioffi, 1980; McDermott, van Horn, and Hansen 1988; Strohmayer, 1991). The models have since been improved to include the following effects: gravitational redshift; the effective boost to the shear modulus due to isotropic magnetic pressure (Duncan, 1998); up to date models of crust composition (Haensel and Pichon, 1994; Piro, 2005); General Relativity and elasticity (Samuelsson and Andersson, 2007). Other factors that need to be assessed include dissipation within the core, viscous effects at the base of the crust, and the presence of crustal inhomogeneities. The effect of the unusual magnetospheric conditions also needs to be taken into account, since the presence of the trapped fireball, which has a major impact on emission properties, will also affect energy loss. For the moment, however, the estimates are such that the modes could survive long enough to explain the observations.

The precise nature of behaviour at the crust/core interface also needs more detailed consideration. Boundary layer physics will be crucial (Kinney and Mendell, 2003), as will the elastic properties at the base of the crust (Pethick and Potekhin, 1998). Coupling will also depend rather sensitively on field geometry: the modes that survive may be those for which coupling to the core is minimal. Indeed the modes found by Glampedakis et al. (2006) had very weak amplitudes in the core. The presence of a strong toroidal field in the core would also increase its rigidity, rendering it less prone to excitation. Until all of these issues are addressed, the global seismic vibration model remains the most promising mechanism for explaining the QPOs, particularly given the difficulties faced by the alternative mechanisms, discussed in the previous section.

## 4.5 Conclusions and Future Work

Neutron stars are the best laboratories for extreme physics in the Universe. Understanding the nature of matter in the cores of neutron stars would be an immense step forward in nuclear physics, and the magnetar QPO detections represent the first serious opportunity to study this region using seismology. Early analysis has already shown the immense potential of this technique to constrain the nuclear equation of state and crust thickness of neutron stars; findings that may rule out strange star models. While there are many theoretical issues still to be resolved, the most urgent requirement is to improve our observational capability so as to obtain the best possible data of future events such as the rare giant flare phenomenon.

In this work, we have presented a detailed analysis of one of the most interesting magnetars, namely SGR 1806-20 and we report the detection of QPOs in the recurrent burst emission. Some of the frequencies we report are within the vicinity of those previously detected from the same source during the giant flare event. We have extended our work to another source (SGR 1900+14), where they also discovered QPOs in the giant flare event. Our results and findings for SGR 1900+14 are forthcoming, where we find strong evidence for the existence of similar oscillations in the recurrent burst emission. Observing QPOs in both the giant flare and the recurrent burst emission from magnetars is an important development in our understanding of this class of exotic compact objects. Further observations, analysis of archival data, and theoretical model developments will be essential in putting more rigid constraints on the neutron star equation of state thus shedding further insight on the phenomena and its interpretation.